\documentclass[useAMS,usenatbib]{mn2e}
\usepackage{graphicx}
\usepackage[figuresright]{rotating}

\def\ngc{NGC\,6231}
\def\hda{HD\,152248}
\def\hd{HD\,152}
\def\cpd{CPD\,$-$41$^{\circ}$}
\def\sco{Sco\,OB\,1}
\def\wr{WR\,79}

\def\l{$\lambda$\,}
\def\ll{$\lambda\lambda$\,}
\def\halph{H$\alpha$}

\def\hbet{H$\beta$}
\def\hea{He\,{\sc i}}
\def\heb{He\,{\sc ii}}
\def\nc{N\,{\sc iii}}
\def\mgb{Mg\,{\sc ii}}
\def\ob{O\,{\sc ii}}
\def\sic{Si\,{\sc iii}}
\def\sid{Si\,{\sc iv}}

\def\kms{km\,s$^{-1}$}

\def\msol{M$_{\odot}$}
\def\s{$\sigma$}

\def\na{{\it n/a}}

\def\sbd{{\sc simbad}}
\def\webda{{\sc webda}}

\defcitealias{SBL98}{SBL98}
\defcitealias{GM01}{GM01}
\defcitealias{HCB74}{HCB74}
\defcitealias{LM83}{LM83}
\defcitealias{PHYB90}{PHYB90}
\defcitealias{PHC91}{PHC91}
\defcitealias{BL95}{BL95}
\defcitealias{RCB97}{RCB97}
\defcitealias{BVF99}{BVF99}
\title[Massive binary fraction in open clusters I. NGC~6231]{The massive star binary fraction in young open clusters I. NGC~6231 revisited}
\author[H. Sana et al.]{H. Sana$^{1}$\thanks{E-mail: hsana@eso.org},
   E. Gosset$^{2}$\thanks{FNRS, Belgium},
   Y. Naz\'e$^{2}$\footnotemark[2],
   G. Rauw$^{2}$\footnotemark[2]
    and N. Linder$^{2}$\thanks{FRIA, Belgium}\\
$^{1}$European Southern Observatory, Alonso de Cordova 1307, Casilla 19001, Santiago 19, Chile\\
$^{2}$Astrophysical Institute, Li\`ege University, Bat. B5c, All\'ee du 6 Ao\^ut 17, B-4000 Li\`ege, Belgium}
\begin{document}

\date{Accepted 1988 December 15. Received 1988 December 14; in original form 1988 October 11}

\pagerange{\pageref{firstpage}--\pageref{lastpage}} \pubyear{2002}

\maketitle

\label{firstpage}

\begin{abstract}
We present the results of a long-term high-resolution spectroscopy campaign on the O-type stars in  NGC 6231. We revise the spectral classification and multiplicity of these objects and we constrain the fundamental properties of the O-star population. Almost three quarters of the O-type stars in the cluster are members of a binary system. The minimum binary fraction is 0.63, with half the O-type binaries having an orbital period of the order of a few days. The eccentricities of all the short-period binaries are revised downward, and henceforth match a normal period-eccentricity distribution. The mass-ratio distribution shows a large preference for O+OB binaries, ruling out the possibility that, in \ngc, the companion of an O-type star is randomly drawn from a standard IMF. Obtained from a complete and homogeneous population of O-type stars, our conclusions provide interesting observational constraints to be confronted with the formation and early-evolution theories of O stars.
\end{abstract}

\begin{keywords}
binaries: close -- binaries: spectroscopic -- binaries: general -- stars: early-type -- 
open clusters and associations: individual: NGC 6231 -- open clusters and associations: Sco OB1
\end{keywords}
%***************************************************************
%*************************************************************** 
\begin{table*}
 \begin{minipage}{170mm}
  \caption{Number of radial velocity (RV) measurements obtained for each object with respect to the considered spectral lines. `--' means that the corresponding line has not been measured. } \label{tab: sp_lines}
 \centering
  \begin{tabular}{@{}rrrrrrrrrrrr@{}}
  \hline
         & \hea       & \sid       & \hea       & \hea       & \hea       & \mgb       & \heb       & \heb       & \hea       & \hea       & \hea     \\
Object   & \l4026     & \l4089     & \l4144     & \l4388     & \l4471     & \l4481     & \l4542     & \l4686     & \l4921     & \l5876     & \l7065   \\
\hline
\hd076   & --         & 1          & 1          & 1          & 1          & 1          & 1          & 1          & 1          & 1          & 1        \\
\hd200   & --         & 16         & 16         & 16         & 16         & --         & 14         & 16         & --         & 16         & 16       \\
\hd233   & --         & --         & 31         & 31         & 34         & --         & 34         & --         & 34         & 31         & 31       \\
\hd234   & 28         & 28         & 28         & 28         & 28         & 28         & 28         & 28         & 28         & 28         & 23       \\
\hd247   & 16         & 16         & 16         & 16         & 16         & --         & 16         & 16         & 16         & 16         & 16       \\
\hd249   & 32         & 34         & 33         & 34         & 34         & --         & 34         & 34         & --         & 29         & 29       \\
\hd314   & 16         & 16         & 16         & --         & 16         & --         & 16         & 16         & 16         & 16         & 16       \\
HD~326329& --         & --         & 8          & 8          & 10         & --         & 8          & 8          & 8          & 8          & 8        \\
HD~326331& --         & --         & --         & --         & 13         & --         & 13         & --         & --         & 13         & --       \\
\cpd7721 & --         & 1          & 1          & 1          & 1          & --         & 1          & 1          & --         & --         & --       \\ 
\hline
\end{tabular}
\end{minipage}
\end{table*}
%***************************************************************
%***************************************************************

\section{Introduction}

Although massive O-type stars are often so bright that they can be studied with small telescopes up to distances of a few kilo-parsecs, our understanding of these objects is still fragmentary. Even their physical properties remain often ill-constrained. The latter however provide key observational guidelines to solve one of the most critical astrophysical questions at present : how do massive stars form ?  In this context, we have undertaken a long-term monitoring of the O-type star population in a number of young open clusters in order to accurately derive and/or confirm their physical and orbital properties. This should help to clarify whether, as suggested by some authors \citep[e.g.\,][]{PGH93, GM01}, a correlation exists between the properties of a cluster and those of its member massive star population. The series of papers initiated here will focus on one of the fundamental properties of these objects, their multiplicity, although other properties of the population will be discussed whenever allowed by the data set. The present paper deals with a supposedly well known cluster, \ngc. Yet, the obtained picture is significantly different from the one proposed in earlier works, both in terms of the properties of the individual objects and those of the O-star population as a whole.

\begin{figure}
\includegraphics[width=\columnwidth]{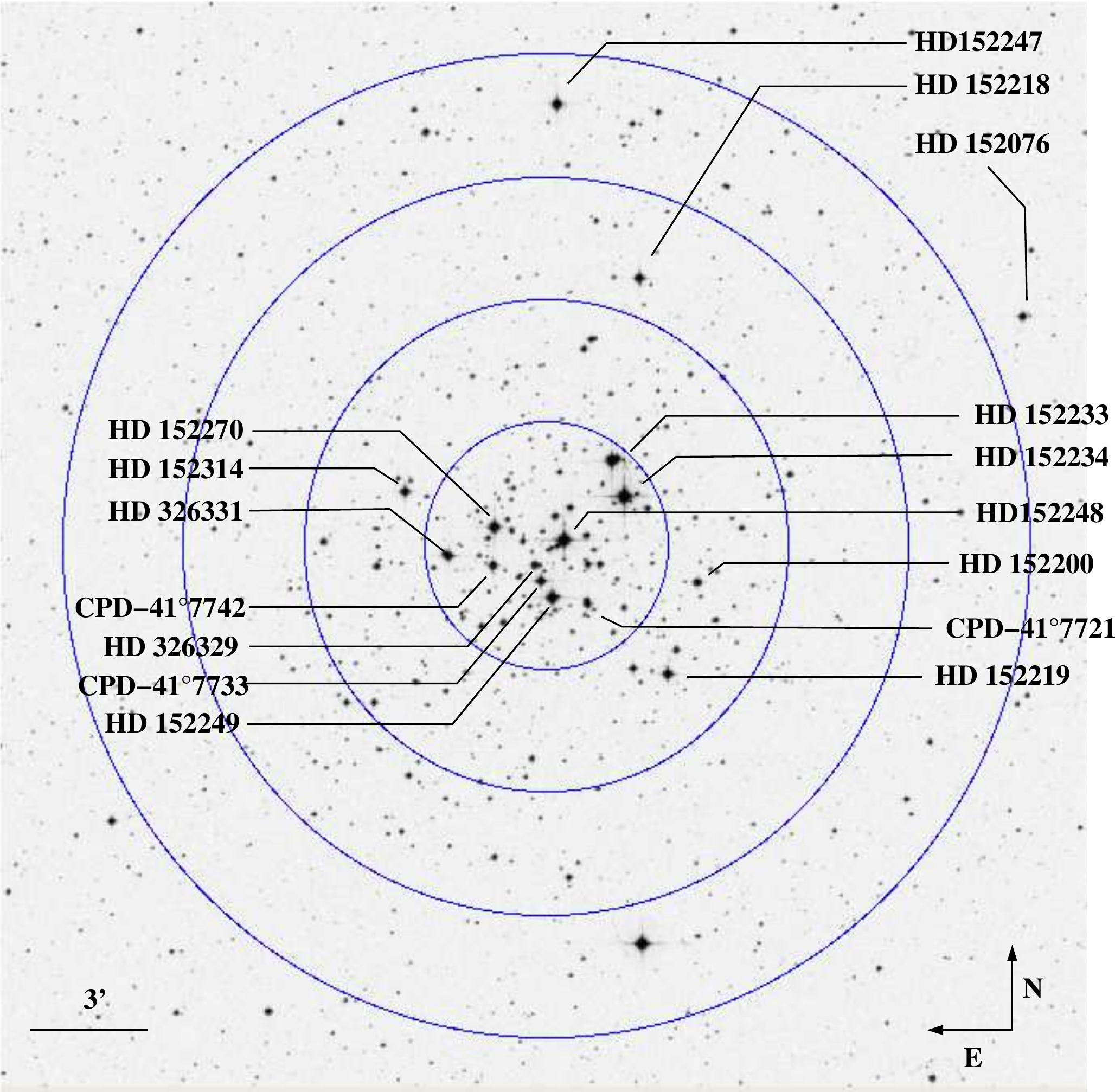}
\caption{O/WR-type objects in  \ngc. The circles are centered on the geometric center of the cluster and have a respective radius of $1\times$, $2\times$, $3\times$ and $4\times$ the cluster core radius $d_\mathrm{c}=3.1\arcmin\equiv 1.5$~pc.  }\label{fig: Oiden}
\end{figure}

Located in the core of the Sco OB1 association, at about 1.64~kpc \citep[$DM=11.07\pm0.04$,][]{SGR06}, \ngc\ is
one of the rich nearby clusters in terms of number of hosted O-type stars \citep[ \citetalias{GM01}]{GM01}. 
\citet{SGR06, SRS07} identified several hundreds of pre-main sequence (PMS) stars, from which they derived a cluster age of 2 to 4~Myr, compatible with the evolutionary status of the massive stars in the cluster, and in good agreement with previous studies \citep{SBL98,BVF99}. Using the X-ray properties of these PMS stars to disentangle the cluster low-mass members from the numerous field stars along the line of sight, they showed that the geometric center of the cluster is located about 30\arcsec\ East of the massive binary \hda\ (Fig.~\ref{fig: Oiden}). Using a King profile fit, they further constrained the cluster core radius at about $d_\mathrm{c}=3.1\arcmin$ which, at the distance of the cluster, corresponds to 1.5~pc.

 Within 15\arcmin\ ($\equiv 7.1$~pc) around the cluster center (about 5 times the core radius), one encounters over 90 B-type stars, 15 O-type stars and a Wolf-Rayet (WR) system \citep[for a complete census, see ][]{SRN06}. The spectroscopic binary (SB) fraction of these objects has already been investigated by several authors. \citet[ \citetalias{LM83}]{LM83} derived a minimum fraction of binaries among main-sequence early-B/late-O stars around $f_\mathrm{min}\sim0.41$, from which the authors estimated a binary fraction around $f\sim0.57$. \citet{Rab96} obtained a minimum fraction $f_\mathrm{min}\sim0.52$ for the B-type stars. More recently, \citetalias{GM01} proposed an even larger fraction of 0.79 for the O-type stars, making \ngc\ the second cluster in their sample, after IC~1805 ($f\sim0.80$), to have an O-star binary fraction close to 80\%. In line with the quoted studies, we adopt here the SB fraction ($f$) to be equal to the ratio of the number of binary or multiple systems to the total number of objects in the considered population. Beyond the SB fraction $f$, the fraction of companions per massive star, as defined by \citet{PWZ01}, will also be briefly discussed in Sect.~\ref{ssect: vis_comp}.

In the present paper, based on more than 150 high signal-to-noise ratio (SNR) high-resolution spectra, we revisited the current knowledge of all the O-type stars in \ngc. Our work is organised as follows. The observing campaign is briefly described in Sect.~\ref{sect: obs}. The individual objects are discussed in Sect.~\ref{sect: objects} while Sect.~\ref{sect: disc} summarizes the properties of the O-type star population seen as a whole. It also discusses the implications in terms of constraints on the star formation process and on the cluster structure. Finally, Sect.~\ref{sect: ccl} briefly summarizes our results.

%***************************************************************
%***************************************************************

%***************************************************************
%***************************************************************
\begin{table*}
 \centering
 \begin{minipage}{120mm}
  \caption{Diagnostic line ratios and adopted spectral classifications for the studied O-type objects. Inappropriate criteria for specific cases are marked as `\na'.} \label{tab: EW}
  \begin{tabular}{@{}lllll@{}}
  \hline
Object        & $\log W'$       &  $\log W''$     &  $\log W'''$       &  Sp. Type\\
  \hline
\hd076        & 0.52            & 0.08 -- 0.09    &  5.3               & O9.5~III  \\
\hd200        & $0.689\pm0.048$ & $0.162\pm0.030$ & $5.438\pm0.022$    & O9.7~V \\
\hd233 prim   & $-0.337\pm0.024$& \na             & \na                & O5.5~III(f)\\
\hd233 sec    & $0.060\pm0.076$ & \na             & \na                & O7.5~III/V?\\
\hd234 prim   & $0.815\pm0.028$ & $0.341\pm0.054$ & \na                & O9.7~I     \\
\hd234 sec    & $\sim0.17$      & $\leq 0.0$      & \na                & O8~V\\
\hd247 prim   & $\sim0.3$       & $\sim0.28$      & \na                & O9~III \\
\hd247 sec    & $ > 0.7$        & $ < 0.0$        & \na                & O9.7~V \\
\hd249        & $0.39\pm0.02$   & $0.554\pm0.004$ &  $4.73\pm0.06$     & O9~Ib~((f)) \\
\hd314 prim   & $0.265\pm0.011$ & $\sim0.27$      & \na                & O8.5~V  \\
\hd314 sec    & \na             & \na             & \na                & B1-3? \\
HD~326329     & $0.49\pm0.04$   & $0.03\pm0.03$   & $5.43\pm0.02$      & O9.5~V \\
HD~326331     & $0.14\pm0.04$   & $0.53\pm0.10$   & \na                & O8~III((f)) \\
\cpd7721      & $\sim0.41$      & $-0.14$ -- $-0.23$ & $\sim5.5$       & O9~V  \\
\hline
\end{tabular}
\end{minipage}
\end{table*}
%***************************************************************
%***************************************************************
%***************************************************************

%***************************************************************
\setcounter{table}{+2}
\begin{sidewaystable*}
\begin{minipage}[t][180mm]{\textwidth} \begin{flushleft}
{\bf Table 3.} Journal of the spectroscopic observations of the \ngc\ O-type stars studied in the present paper. The two header lines indicate the considered spectral lines and the adopted rest wavelength (in \AA). The first column gives the heliocentric Julian date at mid-exposure. The following columns provide the heliocentric RVs (expressed in \kms) using various spectral lines. The last two columns provide the mean and 1-$\sigma$ dispersion computed, for a given date, over the quoted lines. Whenever appropriate, the mean and 1-\s\ dispersion for individual lines are also given at the bottom of each sub-table. References for the instrumental setup can be found at the bottom of the table. The full table is available in the electronic edition of the journal.\end{flushleft}
 \centering
%\label{tab: diary}
\begin{tabular}{@{}rrrrrrrrrrrr@{}}
\hline												
HJD           & \sid\l4089&\hea\l4144&\hea\l4388&\hea\l4471&\heb\l4542&\heb\l4686&\hea\l4921&\hea\l5876 & \hea\l7065 & Mean & Sigma \\
$-2\,400\,000$& 4088.863  & 4143.759 & 4387.928 & 4471.512 & 4541.590 & 4685.682 & 4921.929 & 5875.620  & 7065.190 \\
\hline 
\multicolumn{12}{c}{HD~152200}\\
\hline
51669.900$^b$ &	$-$10.9 &	3.1	 &6.9	 &$-$33.3	 &1.9	 &$-$6.5	&  &$-$6.5	 &$-$5.0		 &$-$6.3	 &12.4 \\
52335.863$^b$ &	$-$31.8 &	$-$19.2	 &$-$16.1	 &$-$58.0	 &$-$29.7	 &$-$32.3	&  &$-$21.8	 &$-$33.0		 &$-$30.2	 &13.0 \\
52336.836$^b$ &	    1.8 &           4.3	 &6.8	 &$-$35.9	 &6.3	 &$-$8.9	&  &$-$1.3	 &$-$7.9		 &$-$4.4	 &14.1 \\
52337.836$^b$ &	$-$10.3 &	$-$0.9	 &6.6 	 &$-$30.5	 &4.8	 &$-$8.5	&  &1.5          &$-$7.4		 &$-$5.6	 &11.9 \\
52338.818$^b$ &	$-$38.4 &	$-$22.7	 &$-$17.4	 &$-$57.4	 &$-$21.3	 &$-$36.9	&  &$-$26.1	 &$-$36.9		 &$-$32.1	 &13.0 \\
52339.806$^b$ &	$-$43.3 &	$-$30.5	 &$-$25.0	 &$-$63.5	 &$-$24.2	 &$-$42.2	&  &$-$39.4	 &$-$37.9		 &$-$38.3	 &12.6 \\
52383.836$^b$ &	$-$46.3 &	$-$33.8	 &$-$32.1	 &$-$69.9	 &$-$14.8	 &$-$36.3	&  &$-$38.2	 &$-$40.4		 &$-$39.0	 &15.5 \\
52782.744$^c$ &	$-$20.9 &	$-$10.8	 &$-$5.3	 &$-$40.6	 &--	 &$-$24.5	&  &$-$13.8	 &$-$18.0		 &$-$19.1	 &11.4 \\
52783.717$^c$ &	$-$46.1 &	$-$35.2	 &$-$26.1	 &$-$63.6	 &--	 &$-$44.7	&  &$-$36.1	 &$-$38.2		 &$-$41.4	 &11.8 \\
52784.715$^c$ &	$-$42.4 &	$-$28.3	 &$-$22.2	 &$-$62.5	 &$-$22.7	 &$-$35.2	&  &$-$30.8	 &$-$39.4		 &$-$35.4	 &13.1 \\
53130.741$^c$ &	$-$58.3 &	$-$47.7	 &$-$43.2	 &$-$79.5	 &$-$41.2	 &$-$58.1	&  &$-$50.5	 &$-$57.0		 &$-$54.4	 &12.1 \\
53131.778$^c$ &	$-$48.6 &	$-$28.1	 &$-$24.2	 &$-$63.8	 &$-$26.1	 &$-$45.8	&  &$-$33.9	 &$-$33.6		 &$-$38.0	 &13.7 \\
53132.645$^c$ &	$-$21.4 &	$-$11.6	 &$-$5.8	 &$-$43.6	 &$-$4.1	 &$-$21.6	&  &$-$17.7	 &$-$17.5		 &$-$17.9	 &12.3 \\
53133.715$^c$ &	$-$26.4 &	$-$12.0	 &$-$8.2	 &$-$45.2	 &$-$5.5	 &$-$26.4	&  &$-$15.9	 &$-$20.8		 &$-$20.1	 &12.8 \\ 
53134.827$^c$ &	$-$55.0 &	$-$42.0	 &$-$35.4	 &$-$74.4	 &$-$35.1	 &$-$55.3	&  &$-$45.0	 &$-$47.9		 &$-$48.8	 &12.9 \\
53135.676$^c$ &	$-$54.7 &	$-$43.6	 &$-$39.3	 &$-$77.5	 &$-$35.1	 &$-$59.1	&  &$-$50.0	 &$-$52.8		 &$-$51.5	 &13.2 \\
Mean	      &	$-$34.6 &	$-$22.4    &$-$17.5  &$-$56.2  &$-$17.6  &$-$33.9 &  &$-$26.6        &$-$30.9          &$-$30.2  &12.9 \\		
Sigma	      &  18.1 &	16.5     &16.3   &16.0   &15.8   &17.0  &  &16.6         &16.3           &16.3   &1.0  \\	
%\hline
%\multicolumn{12}{c}{CPD$$-$41$\degr7721}\\
%\hline									
%53134.767$^c$ & $-$26.8 &	$-$13.4    &$-$15.2  &	$-$35.9&	$-$36.5&	$-$25.3&$-$12.4	&$-$18.1 &$-$18.2 &	$-$22.4&	9.2 \\
\hline					
\multicolumn{12}{c}{HD~326329 $\equiv$ \cpd7735} \\
\hline													
50995.687$^a$   & &    --    &   --  & $-$38.7 &   --  & --    & --    & --   & --       & --           & --    \\
50996.617$^a$   & &    --    &   --  & $-$36.7 &   --  & --    & --    & --   & --       & --           & --    \\
51671.911$^b$   & &    $-$20.0 & $-$18.3 & $-$19.6 & $-$38.1 & $-$21.5 & $-$24.7 & $-$18.1&	$-$14.6  & 	$-$21.86&	 7.2  \\
51672.901$^b$   & &    $-$14.9 & $-$17.5 & $-$26.2 & $-$33.5 & $-$21.9 & $-$19.9 & $-$14.5&	$-$13.7  & 	$-$20.26&	 6.8  \\
53130.709$^b$   & &    $-$15.9 & $-$20.5 & $-$28.2 & $-$22.0 & $-$18.1 & $-$22.2 & $-$17.8&	$-$14.9  & 	$-$19.95&	 4.3  \\
53131.747$^b$   & &    $-$15.0 & $-$20.6 & $-$28.6 & $-$22.7 & $-$18.6 & $-$23.5 & $-$16.8&	$-$15.6  & 	$-$20.18&	 4.6  \\
53132.781$^b$   & &    $-$18.0 & $-$21.6 & $-$31.4 & $-$27.7 & $-$21.6 & $-$25.2 & $-$20.7&	$-$19.3  & 	$-$23.19&	 4.6  \\
53133.816$^b$   & &    $-$16.8 & $-$20.9 & $-$26.7 & $-$22.7 & $-$20.3 & $-$23.3 & $-$17.6&	$-$19.5  & 	$-$20.98&	 3.2  \\
53134.691$^b$   & &    $-$17.7 & $-$20.8 & $-$29.5 & $-$23.6 & $-$18.0 & $-$24.7 & $-$19.0&	$-$18.1  & 	$-$21.43&	 4.2  \\
53135.735$^b$   & &    $-$18.1 & $-$21.8 & $-$29.9 & $-$23.4 & $-$23.2 & $-$23.9 & $-$19.5&	$-$16.6  & 	$-$22.05&	 4.1  \\   
Mean	        & &    $-$17.1 & $-$20.3 & $-$29.6 & $-$26.7 & $-$20.4 & $-$23.4 & $-$18.0&	$-$16.5  & 	$-$21.2 &  4.9  \\
Sigma 	        & &      1.7 &   1.5 &   5.4 & 6.0   & 2.0   & 1.7   & 1.9  &     2.2  &          1.1 &  1.4  \\
\hline
\end{tabular}\\
\begin{flushleft}$^a$ CAT+CES ; $^b$ ESO1.5m + FEROS ;   $^c$ ESO/MPG2.2m + FEROS ;   $^d$ CTIO1.5m + BME. 
\end{flushleft}
\end{minipage}
\end{sidewaystable*}
%***************************************************************
%***************************************************************
%***************************************************************

\section{Observations and data handling}\label{sect: obs}

Our team started to collect data on the O-type stars in \ngc\ about 10 years ago, focusing in the first few years on the brightest objects before later extending our survey to  fainter targets in the cluster. Except for a few spectra obtained with the Coud\'e Echelle Spectrograph (CES) at the ESO Coud\'e Auxiliary Telescope (CAT, La Silla) in 1998, and with the Bench-Mounted Echelle spectrograph (BME) attached to the CTIO 1.5m Ritchey-Chr\'etien Telescope at Cerro Tololo, most of the present data set was acquired between May 1999 and May 2004 with FEROS (Fiber-fed Extended Range Optical Spectrograph) successively  mounted at the ESO-1.5m and ESO/MPG-2.2m telescopes at La Silla.  The data reduction techniques applied are identical to those used in previously published analyses of the bright short-period binaries in the cluster. We refer to our previous works on these objects for a complete description of the instrumental setups and of the reduction techniques \citep[see e.g.\,][]{SRG01, SHRG03, SRG07}.
For each spectrum, Doppler shifts and equivalent widths (EWs) of a series of lines (Table \ref{tab: sp_lines}) were measured by fitting Gaussian profiles to the spectral lines. For this purpose, we adopted the effective rest wavelengths from \citet{CLL77} below 4800 \AA\ and from \citet{Und94} above that limit. % For the metallic lines that are not listed in these latter works, we used the rest wavelengths from \citet{Moore59}.
 The spectral properties of the O-type stars were re-derived using the quantitative criteria of \citet{CA71}, \citet{Con73_teff}, \citet{Mat88} and \citet{Mat89}, that  rely on the EW ratio of diagnostic lines. We adopt the usual notations: $\log W'= \log W(\lambda4471) - \log W(\lambda4542)$,  $\log W''= \log W(\lambda4089) - \log W(\lambda4144)$ and  $\log W'''= \log W(\lambda4388) + \log W(\lambda4686)$. With respect to the latter criterion, based on the product of the EWs (expressed in m\AA), we emphasize that the star has to be single, or that the brightness ratio of the two components has to be known for this criterion to be applicable.  Measured values are reported in Table \ref{tab: EW}. Whenever possible, B-type companions were also classified by comparing the measured EWs with the typical EWs, quoted by \citet{Did82}, for a given spectral type. 

Table 3 summarizes the derived radial velocities (RVs)  and provides, object by object, the journal of the observations of all the O-type stars, with the exception of the 5 short-period binaries, that have been the subject of dedicated papers, and of the WR system. References for the latter objects are given in Table ~\ref{tab: shortP}. Tables \ref{tab: sp_lines} and 3 show that, for most objects, we are able to investigate variability on time scales of several hours to several years.

To further study the possible long-term variability, we also compared our measurements with results from earlier works. In the past, RV measurements on cluster stars have mainly been performed by \citet{Str44}, \citet[ \citetalias{HCB74}]{HCB74}, \citet[ \citetalias{LM83}]{LM83}, \citet[ \citetalias{PHYB90}]{PHYB90} and \citet[ \citetalias{GM01}]{GM01}. Note however that these authors report (mean) RVs based on different lines, so that direct comparison of the quoted RVs from one author to the other, or with the current work, should be considered as indicative only. The most recently published photometric studies are from  \citet[ \citetalias{PHC91}]{PHC91}, \citet[ \citetalias{BL95}]{BL95}, \citet[ \citetalias{RCB97}]{RCB97}, \citet[ \citetalias{SBL98}]{SBL98} and  \citet[ \citetalias{BVF99}]{BVF99}. Sung (2005, private communication) also obtained UBV(RI)$_\mathrm{C}$ photometry over a 40\arcmin\ $\times$ 40\arcmin\ field of view down to $V\sim23$ \citep[for a brief description of the data, see ][]{SGR06}.

Investigating the multiplicity of massive stars is a difficult task, not only because of the numerous observational biases, but also because of the large parameter space that must be searched. In the present approach, we adopt either of the following two criteria as a definite proof that an object is a binary system: (i) the presence of Keplerian (and periodic) RV variations, or (ii) the detection of a composite (SB2) and variable spectrum. 

In the first case, we are able to compute an orbital solution, thus constraining most of the orbital parameters. In the second case, the detection of the secondary spectrum also provides valuable information. Indeed, the spectral types of both components offer a reasonable estimate of the mass ratio and the detected RV variations yield a first, albeit rough, idea of the time scale of the orbital period. 

As described below, some objects in our sample show RV variations that we have not been able to link with a Keplerian motion, nor have we been able to detect the secondary signature. While some of these objects are potential binary candidates, their multiplicity nevertheless awaits more definite confirmation. Combined with the fact that we are actually observing the entire massive star population in the cluster and not a sub-sample of it,  the present approach offers the advantage of providing a firm lower limit on the binary fraction in \ngc, which is unaffected by statistical errors.

\section{The O-type star population in NGC~6231} \label{sect: objects}

\setcounter{table}{+3}
\begin{table*}
 \centering
 \begin{minipage}{160mm}
  \caption{Selected orbital and physical parameters of the short period binaries in \ngc.} \label{tab: shortP}
  \begin{tabular}{@{}llllll@{}}
  \hline
Object & $P$ (d) & $e$ &  $m_1/m_2$ & Sp. Type & Reference \\
 \hline
\cpd7742  & $2.44070$ & $0.027^a\pm0.006$ & $1.803\pm0.015$ & O9~V + B1.5~V             & \citet{SHRG03}\\
\hd219    & $4.24032$ & $0.082\pm0.011$   & $2.530\pm0.023$ & O9~III + B1-2~V/III       & \citet{SGR06_219}\\
\hd248$^b$& $4.81602$ & $0.133\pm0.005$   & $1.012\pm0.011$ & O7~III~(f) + O7.5~III~(f) & \citet{SRG01}\\
\hd218    & $5.60391$ & $0.259\pm0.006$   & $1.319\pm0.014$ & O9~IV + O9.7~V            & \citet{SNoD07}\\
\cpd7733  & $5.681504$&  0.0 (fixed)      & $2.640\pm0.012$ & O8.5~V + B3               & \citet{SRG07} \\
\wr (\hd270)&$8.8911$ &  0.0 (fixed)      & 2.7             & WC7 + O6V                 & \citet{Lue97,HMSL00}\\
\hline
\end{tabular}\\
$^a$ We note that \citet{BSP05} reported a circular orbit for \cpd7742.\\
$^b$ The orbit has been recomputed using the new version of the Li\`ege orbital solution package \citep{SGR06_219} for a better handling of the errors.
\end{minipage}
\end{table*}

As mentioned earlier, 15 O-type objects are found within 5 core radii from the cluster center (Fig.~\ref{fig: Oiden}). In the present section, we revisit the properties of each of these objects individually.

\subsection{Short period binaries} \label{ssect: shortP}
Detailed analyses of the 5 short period binaries were presented in a series of papers since 2001. Table \ref{tab: shortP} summarizes the main physical and orbital properties of these objects. We note that they significantly differ from the results quoted in the contemporaneous work of \citetalias{GM01}. Implications for the general properties of the \ngc\ O-star population, as well as constraints on our understanding of the formation and dynamical evolution of these objects,  will be discussed in Sect.~\ref{sect: disc}. 

\subsection{Long period binaries} \label{ssect: longP}
%***********************************************
\begin{figure}
\centering
\includegraphics[height=8.0cm]{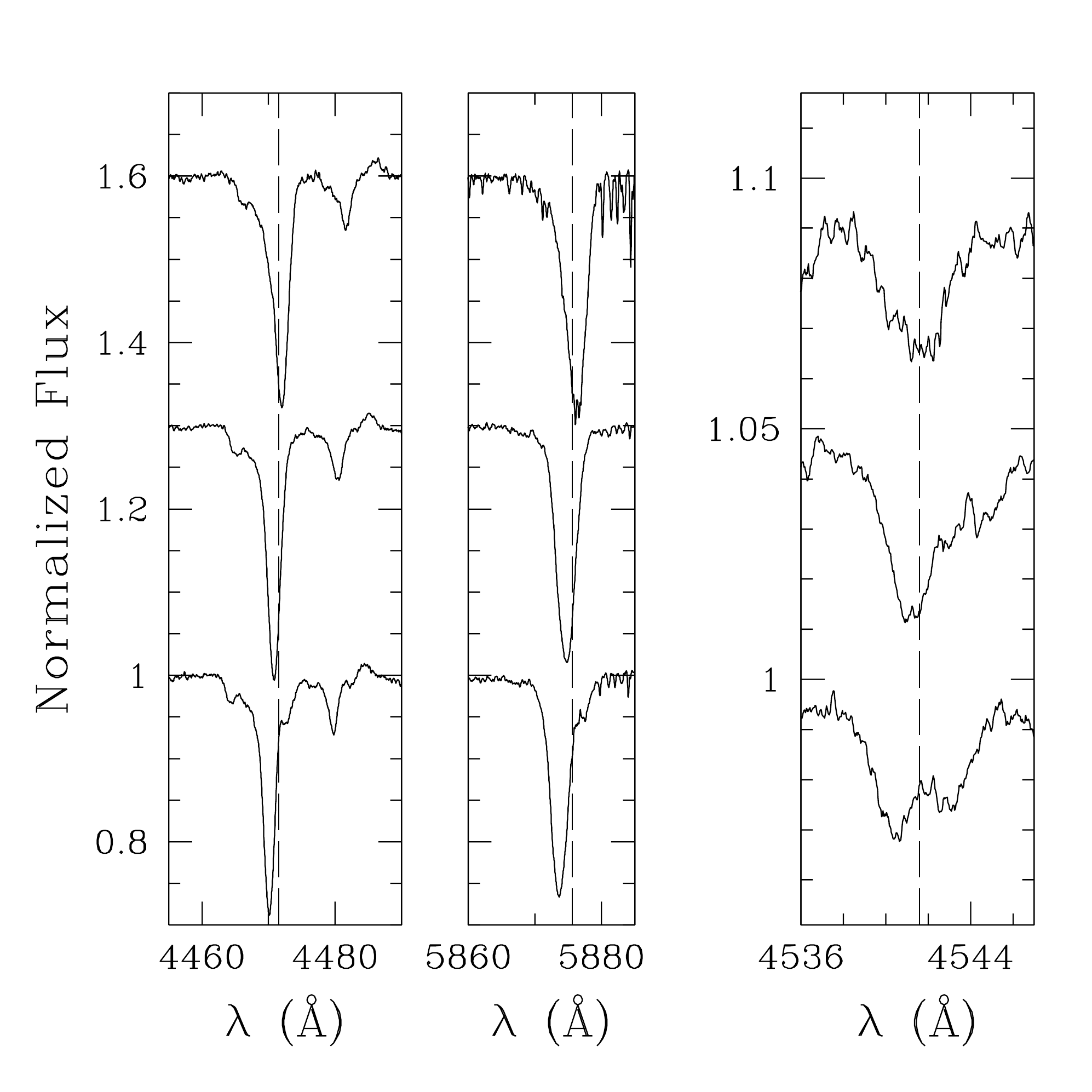}
\caption{{\bf \hd234:} \hea\,\ll4471, 5876 (left) and \heb\,\l4542 (right) line profiles at three different epochs. The orbital shift of the primary is clearly seen. The secondary component is clearly visible, blueshifted, in the top spectrum  and redshifted in the  bottom one. Vertical dashed lines indicate the adopted rest wavelengths.}
\label{fig: hd234}
\end{figure}
%***********************************************

\subsubsection{HD~152234} \label{ssect: hd234} 

\hd234 is reported in the literature as a B0Iab  radial velocity variable star \citep{LM80}. \citet{LMGM88}  proposed a very preliminary period of 27.25 days, while  \citetalias{GM01} claimed to have found the companion signature and derived a period twice as long ($P=54.6$~d). Their computed orbital solution is slightly eccentric ($e=0.183\pm0.093$) and they estimated $K_1=46.5\pm2.8$~\kms\ and $K_2=211.5\pm6.1$~\kms (r.m.s.~=~13.5~\kms) . \\

\begin{figure}
\centering
\includegraphics[width=\columnwidth]{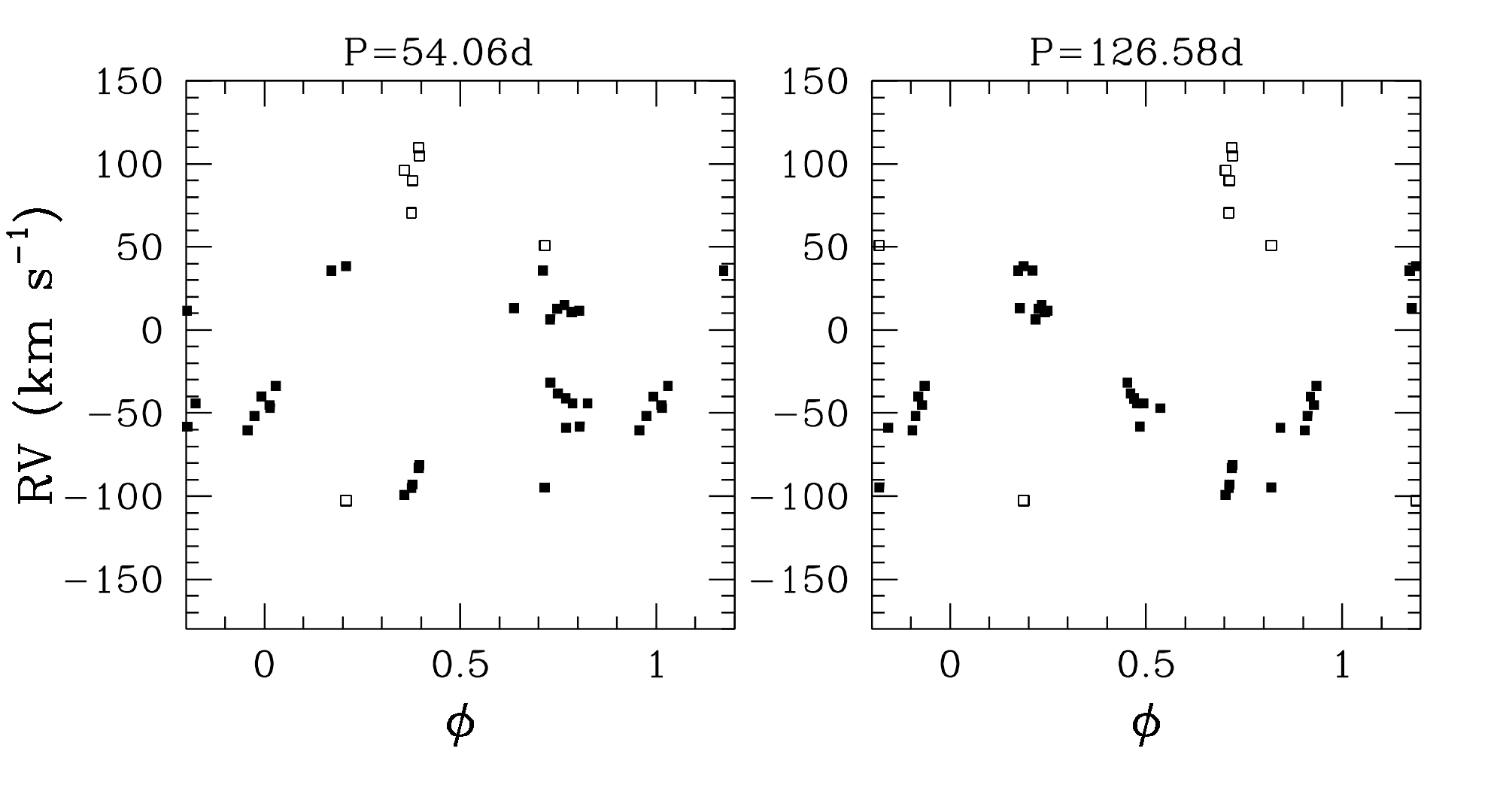}
\caption{{\bf \hd234:} \hea\l4471 RVs folded against two values of the orbital period. Filled (resp. open) symbols indicate primary (resp. secondary) RVs.}
\label{fig: hd234_orbit}
\end{figure}

Between 1999 and 2004, we have obtained 23 FEROS spectra and 5 BME spectra.  From our observations, HD 152234 most probably consists of an O9.7 supergiant primary and an O8 main sequence secondary. The secondary signature is faint (Fig.\,\ref{fig: hd234}) and is only seen at very few phases. This partly results from the lower SNR of some of the spectra in our data set.  Our current data definitely rule out the period of 54.6~d proposed by \citetalias{GM01} and point towards a preliminary value close to 126~d (Fig.~\ref{fig: hd234_orbit}). Although the orbital properties await further constraints, we nevertheless confirm that \hd234 is a long period O-type binary.  \\

%***********************************************
%***********************************************

%***********************************************
\begin{figure}
\centering
\includegraphics[width=\columnwidth]{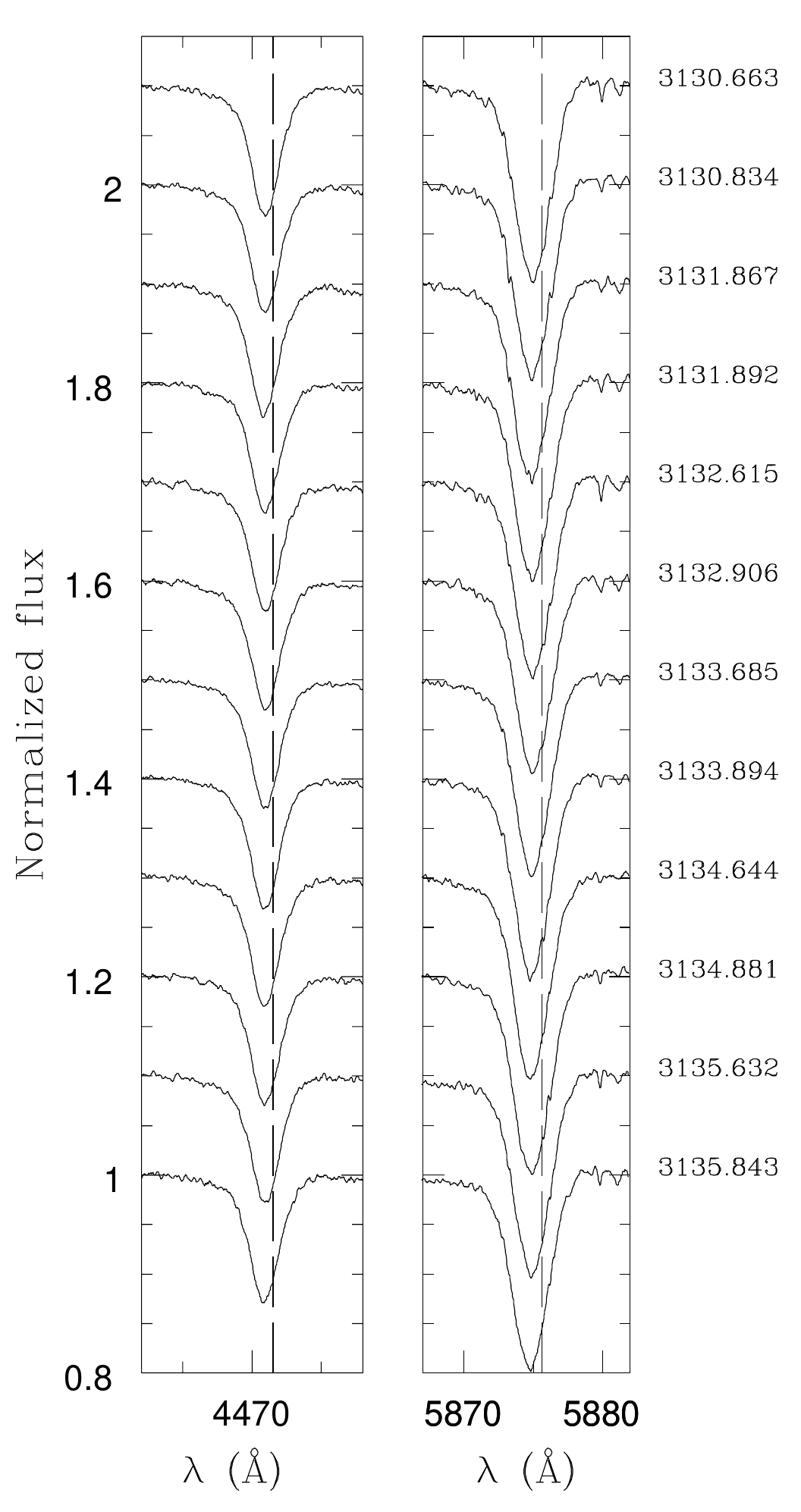}
\caption{{\bf \hd233:} \hea\,\ll4471 and 5876 line profiles as observed during a 6-night run in May 2004. HJD at mid-exposures are quoted at right-hand in the format HJD$-$2~450~000. The vertical dashed lines represent the rest wavelength. No significant difference in the position of the absorption lines is observed from one spectrum to the other, ruling out the 4.15~d period derived by \citetalias{GM01}. }
\label{fig: hd233a} 
\end{figure}
%***********************************************

%***********************************************
\begin{figure}
\centering
\includegraphics[width=\columnwidth]{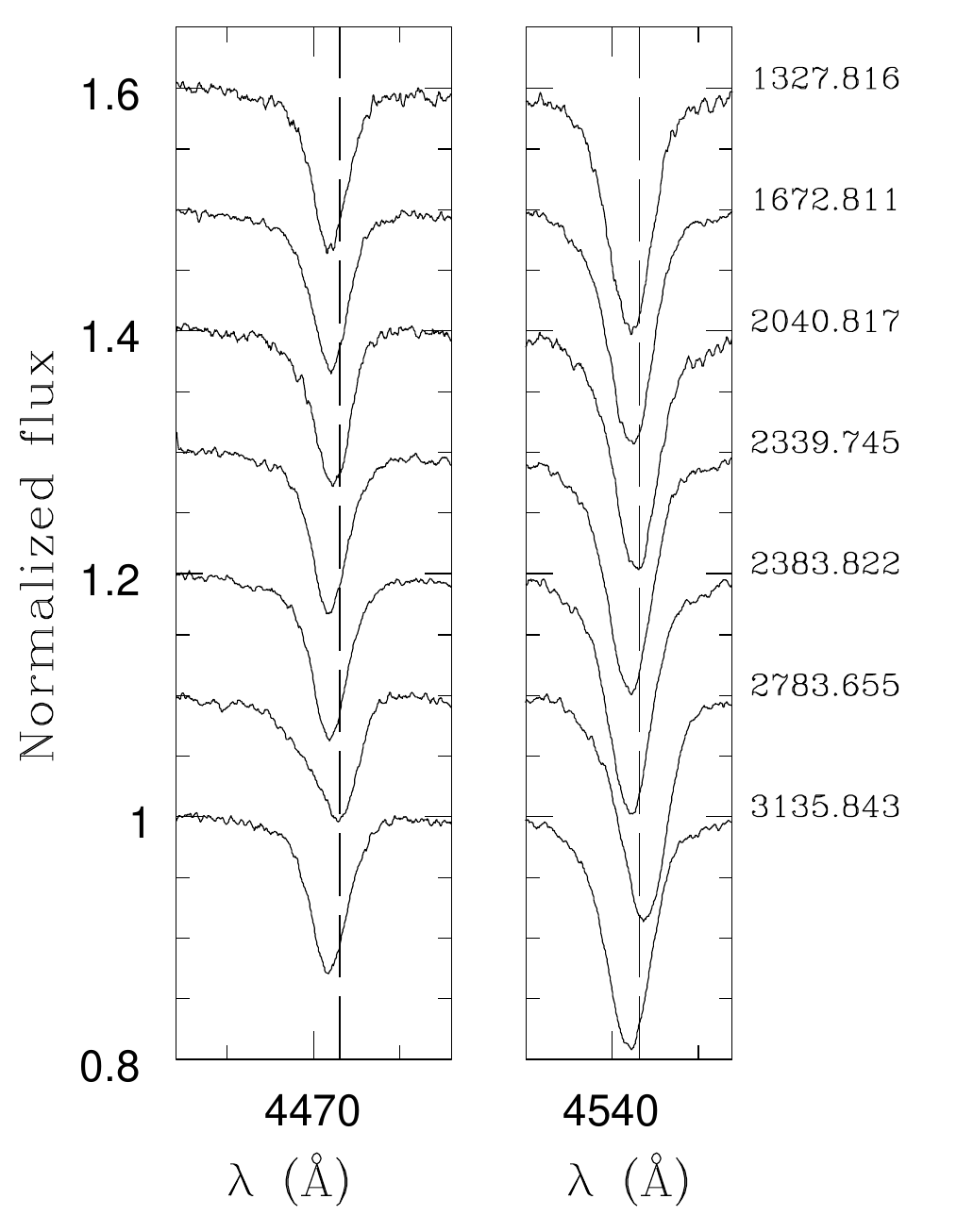}
\caption{{\bf \hd233:} \hea\,\l4471 and \heb\,\l4542 line profiles as observed during different observing runs from 1999 to 2004.  Clear shifts are seen and the profile asymmetry in May 2003 (HJD$\sim$2~452~784) suggests  that the secondary is a late O-type star. The layout of the figure is similar to the one of Fig.~\ref{fig: hd233a}. }
\label{fig: hd233b} 
\end{figure}
%***********************************************

\subsubsection{HD~152233} \label{ssect: hd233}
\hd233 is known as an O6III(f) star \citep{Wal72, LM80} that displays small RV shifts \citep[e.g.\,][]{Str44}. Based on data from the literature along with four additional  observations, \citetalias{GM01} proposed a short period of 4.15~d. According to their Table 2, they  measured the same RV in two spectra separated by one day. Consequently, they adopted an extremely large eccentricity $e=0.57$, thus making \hd233 a potential cornerstone for the study of the dynamical evolution of O-type binaries. \\

 In May 2004, we observed the system twice a night for six consecutive nights, thus well sampling the proposed 4.15~d period. However, we definitely ruled the latter out as we only detected a slow trend in the system RVs (Fig. \ref{fig: hd233a}). The current data set points to a period of a few hundred days but awaits further constraints. Our May 2003 data revealed for the first time the signature of the secondary component in the blue wing of the \hea\ and \heb\ lines (Fig. \ref{fig: hd233b}). Despite the fact that the lines are only marginaly disentangled, we estimated individual spectral types of O5.5 and O7.5 for the two components. The \heb\l4686 line displays a variable P-cygni profile.  Although the current data set does not allow us to separate the relative contributions of the two components to the \heb\l4686 profile, its EW  is compatible with both stars being giants.   Finally, we also detected \nc\ lines that are usually not present in a typical O5.5~III spectrum. These lines are  moving in the opposite direction compared to the primary lines. Clearly this is also the signature of the companion,  a less evolved O star of spectral type O7.5 or later. While more numerous disentangled spectra are needed to better constrain the nature of the secondary, \hd233 is definitely a long period O+O binary.\\

\subsubsection{\hd247} \label{sect: 247}

At 11\arcmin\ N of \hda, \hd247 has clearly suffered from a lack of attention in the previous works. Mostly classified so far as O9.5~III \citep{Ho56, FF68, Mat88, PGB96}, the star seems to be constant in the $V$ band \citep{Bal83} with a range in the published values smaller than 0.05 mag. Owing to the absence of the star in the usual photometric works on the cluster,  we adopted the magnitude of \citet{DiS94}: $V=7.18$. The first RVs were obtained by \citet{Str44} who quoted values around $-16$~\kms (rms$\sim$10~\kms). No other measurement was obtained until \citet{Rab96} suggested that the star was displaying RV shifts. He indeed reported 5 measurements spread over two observing epochs (see the \webda\ database\footnote{http://www.univie.ac.at/webda}). The first set was acquired over three consecutive nights and indicated values around +20($\pm$3)~\kms. The next two measurements were obtained about two years later, again on two consecutive nights, and yielded values close to $-23$($\pm$2)~\kms. More recently, \citet{SL01} reported two additional observations from the International Ultraviolet Explorer (IUE) obtained about 10 years before \citet{Rab96} and indicating $RV\sim-39$($\pm$1)~\kms. Though the star clearly showed RV variations, the time scale over which these occur remained very poorly constrained.

%***********************************************
\begin{figure*}
\includegraphics[width=\textwidth]{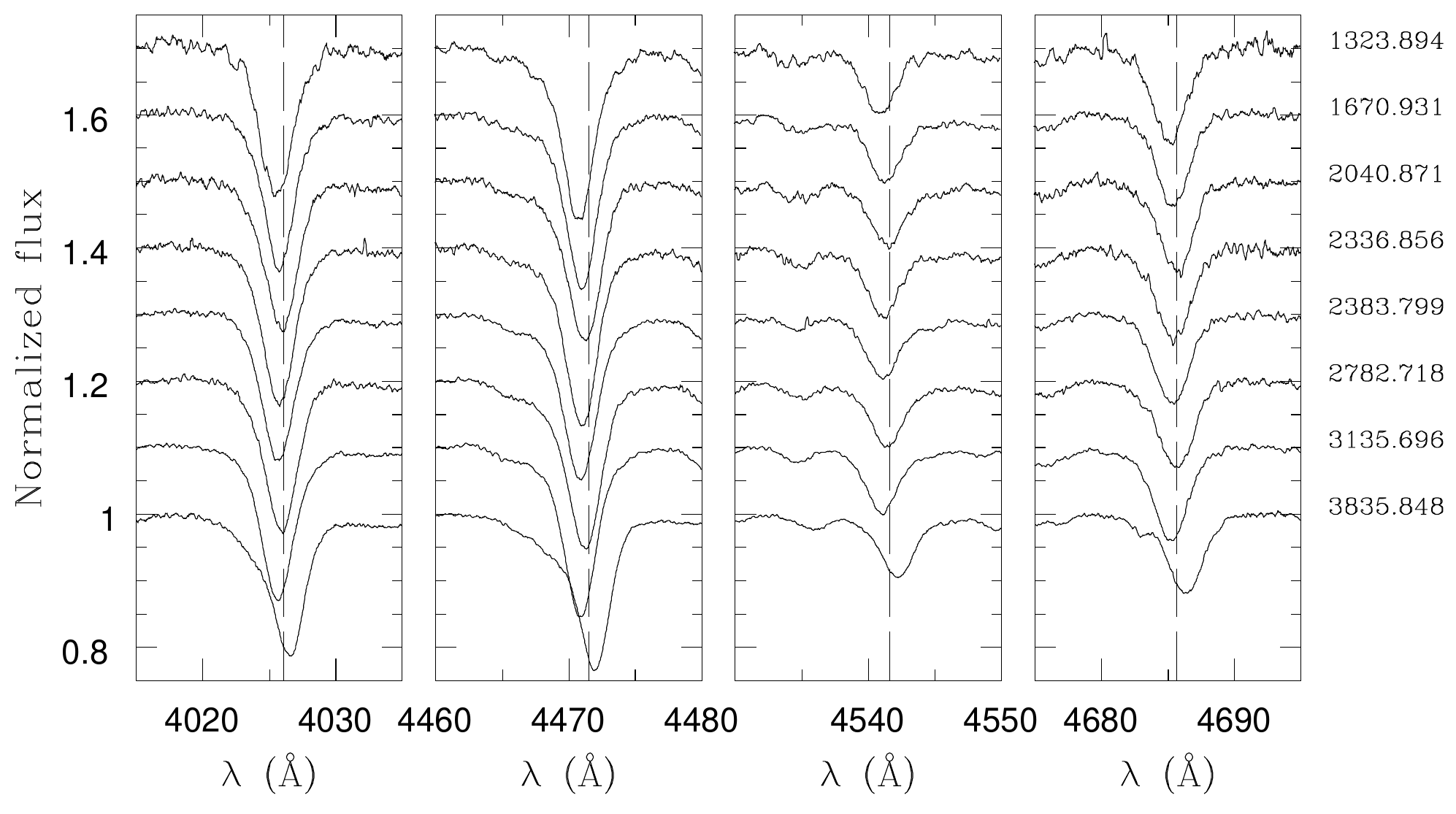}
\caption{{\bf \hd247:} \hea\,\ll4026, 4471 and \heb\,\ll4542, 4686 line profiles as observed during different runs from 1999 to 2006. The spectral signature of the companion is clearly seen in the 2006 spectrum (at HJD$\sim$2~453~836).  The layout of the figure is similar to the one of Fig.~\ref{fig: hd233a}.}
\label{fig: hd247a}
 \end{figure*}
%***********************************************

%***********************************************
\begin{figure*} \centering
\includegraphics[width=.8\textwidth]{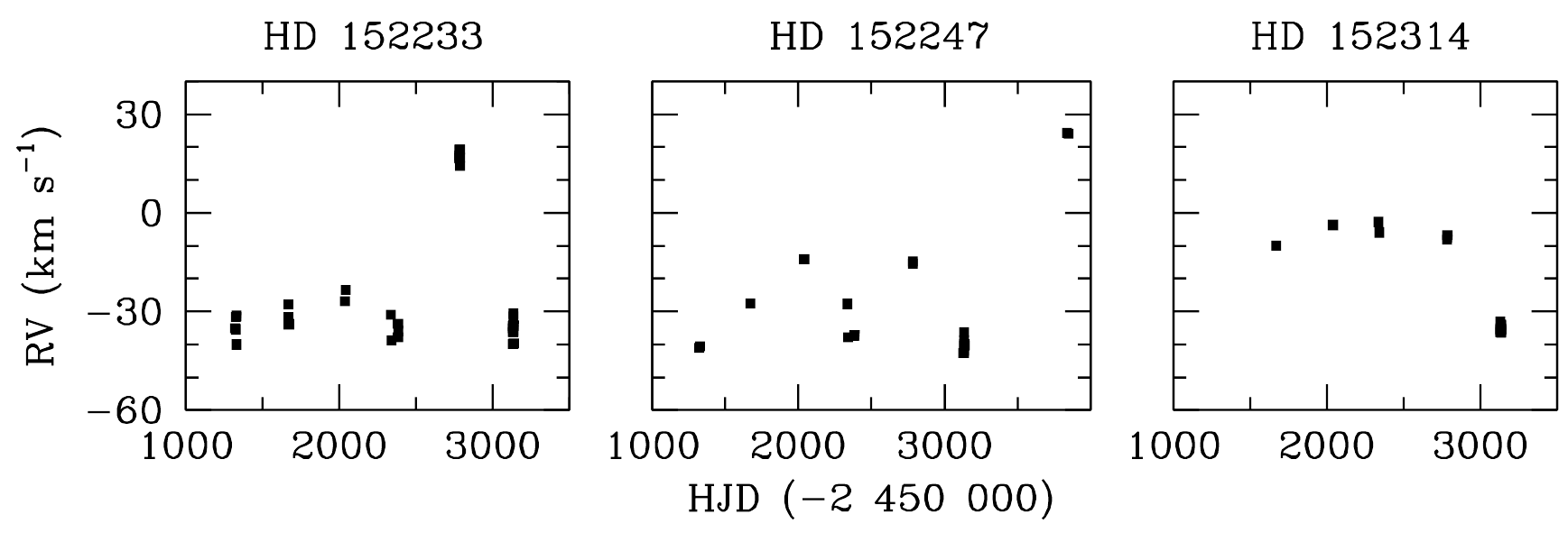}
\caption{RV variations of the three long period binaries with ill-constrained periods: \hd233 (\hea\l5876), \hd247 (\hea\l5876) and \hd314 (\heb\l4686). }
\label{fig: longP}
\end{figure*}
%***********************************************
In our data, clear RV shifts are noticeable from year-to-year (Figs.~\ref{fig: hd247a} and \ref{fig: longP}).  The two extreme RV values in our set are approximately $-40$ (1999, 2004) and $+20$~\kms\ (2006). In May 2004, we obtained six spectra over six consecutive nights but the \hd247 spectrum remained almost unchanged. Similarly no significant shift was observed between the 2002 March and April data. In April 2006, we unambiguously detected for the first time the signature of the secondary star. Values for $\log W'$ and $\log W''$ point towards the primary being an O9\,III star. 
In addition, the secondary displays a clear \heb\l4686 line in absorption and is thus an O-type star as well. Because of the blending with neighbouring primary lines, the disentangling of the secondary classification lines is difficult. To correct for the primary line, we subtracted a spectrum where the two components were deblended from one where they were blended, taking into account the measured Doppler shift. In principle, the primary signature should be removed, leaving a positive and negative image of the secondary spectrum shifted from one another. Using this rough method, we estimate that the secondary is probably an O9.7 dwarf.

Again, more disentangled spectra are needed to confirm the  classification of the secondary and to constrain the orbital period, which is probably of the order of years. Still, the present results indicate that \hd247 is an additional O+O binary in \ngc.\\

\begin{figure*} \centering
\includegraphics[bb=0 0 573 383,width=\textwidth,clip]{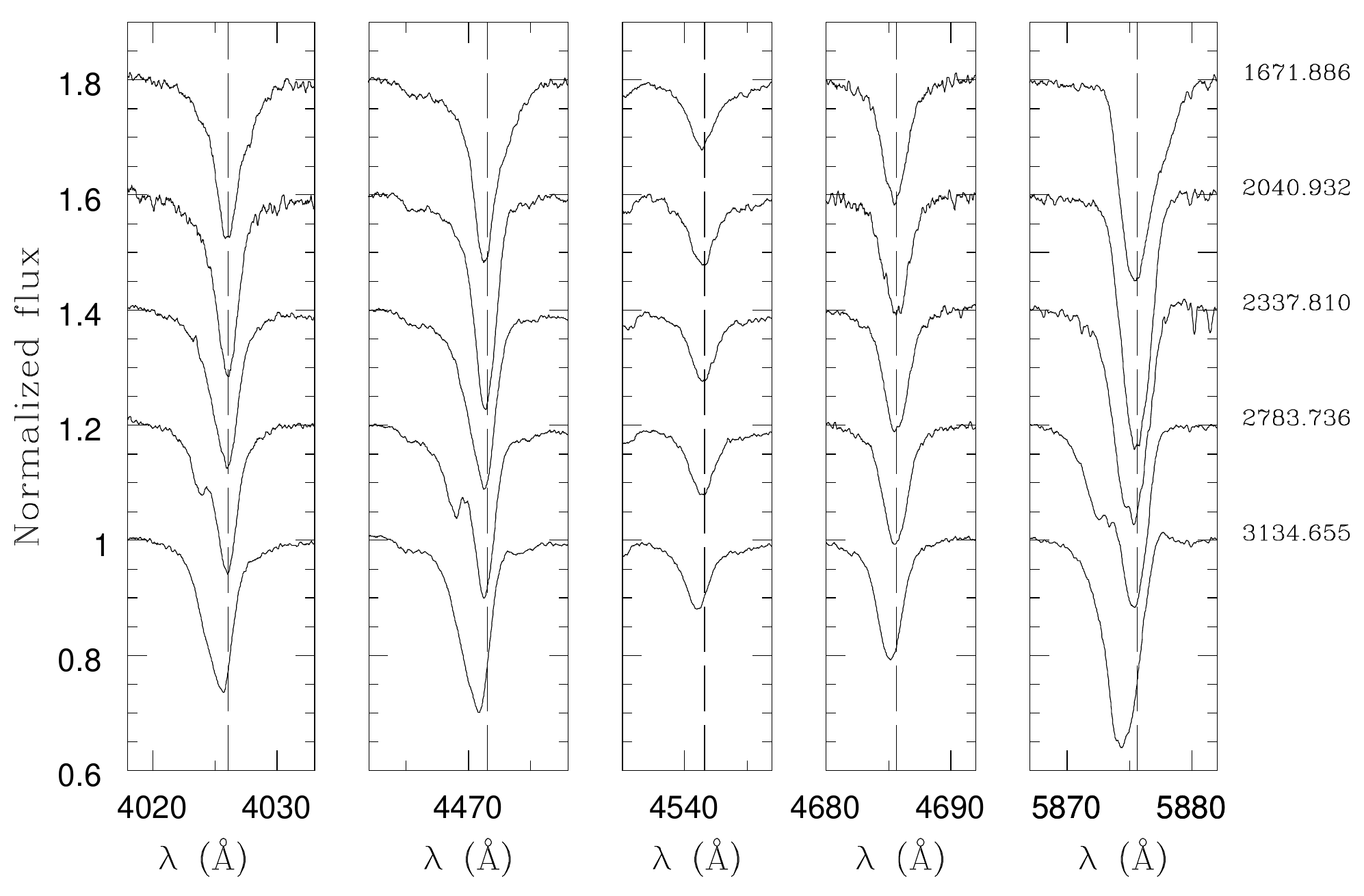}
\caption{{\bf \hd314:}  \hea\,\ll4026, 4471, \heb\,\ll4542, 4686 and \hea\,\l5876  lines acquired during five different observing runs.  The spectral signature of the companion is clearly seen in the May 2003 spectrum (at HJD$\sim$2~452~784).  The layout of the figure is similar to the one of Fig.~\ref{fig: hd233a}.}
\label{fig: hd314b}
\end{figure*}

\subsubsection{\hd314} \label{sect: 314}

At 4\farcm2 E-NE of the cluster core, \hd314 is a late O-type object.  With  $V=7.92-8.12$, \citet{FF68} reported that the star displays photometric variability. \citetalias{RCB97} further found long-term variability on a time-scale of $\sim18$ years. These variations display different amplitudes in the different bands of the Geneva system: the star is becoming bluer when it brightens. \citetalias{RCB97} also suggested an additional, shorter time-scale variability of the order of 1.5 months. RV measurements by \citet{Str44}, \citetalias{PHYB90} and \citetalias{LM83} do not present any particular variation and the latter authors reported the star as RV constant. More recently, \citetalias{GM01} obtained three additional spectra: two separated by 4 days,  another acquired about two years later. Though with some uncertainty on the RV measurements, they reported to have detected the signature of a blue-shifted secondary component on the first spectrum of their series, thus suggesting variability on a time-scale of days. The spectral types  quoted in the literature are O8.5\,III \citep{LM80}, O9~III \citep{SHS69, PHYB90} or O9~V \citep{MGG53}. The remark by \citetalias{RCB97} that the object's colour is changing with time could also indicate a slight change in its spectral type. 

We have acquired a few FEROS spectra of \hd314 since May 2000. In May 2004, we acquired 9 spectra over a 6-night run. The night-to-night variability is very limited but we observed significant changes from one year to the other (Figs.~\ref{fig: longP} and \ref{fig: hd314b}). In May 2003, we detected the presence of a secondary companion in most of the \hea\ and Balmer lines.  The \halph\ profile presents mixed absorption and emission and evolves from an inverse P-Cygni profile to a normal P-Cygni through the years, a behaviour that could be compatible with an absorption component moving on top of a slightly broader emission.  
The spectral type changes from O9 to O8.5 between 2000 and 2004, but this could be due to an effect of line blending. The best spectral type estimate is probably obtained in 2003, when the two components were deblended, which indicates an O8.5~III primary. The \heb\ lines seem to belong solely to the primary. The secondary component is thus probably a B-type star. However, its signature remains undetected in the metallic lines, rendering an accurate spectral classification very difficult. Because of the strong secondary signature in the \hea\ lines, its classification could be located around the \hea\ maximum, i.e.\ at spectral-types B1--B3. This classification remains however uncertain. Finally, with $V=7.749$ and $B-V=0.525$, we obtained $M_\mathrm{V}=-4.95$ for the system, significantly fainter than a typical O8.5~III star \citep[$M_\mathrm{V}^\mathrm{O8.5 III}=-5.32$,][]{MSH05}. This situation is reminiscent of the one observed for other O-type binaries in the cluster. We thus assume that the primary is rather a dwarf and we adopt, as our best classification, O8.5~V + B1-3~V.
 This object definitively deserves follow-up observations covering time scales from weeks to years.\\

%%%%%%%%%%%%%%%%%%%%%%%%%%%%%%%%%%%%%%%%%%%%%%%%%%%%%%%%%%%%%%%%%%%

\subsection{Variable stars} \label{ssect: var}

%***********************************************
\begin{figure*} \centering
\includegraphics[bb=0 0 573 236,width=\textwidth,clip]{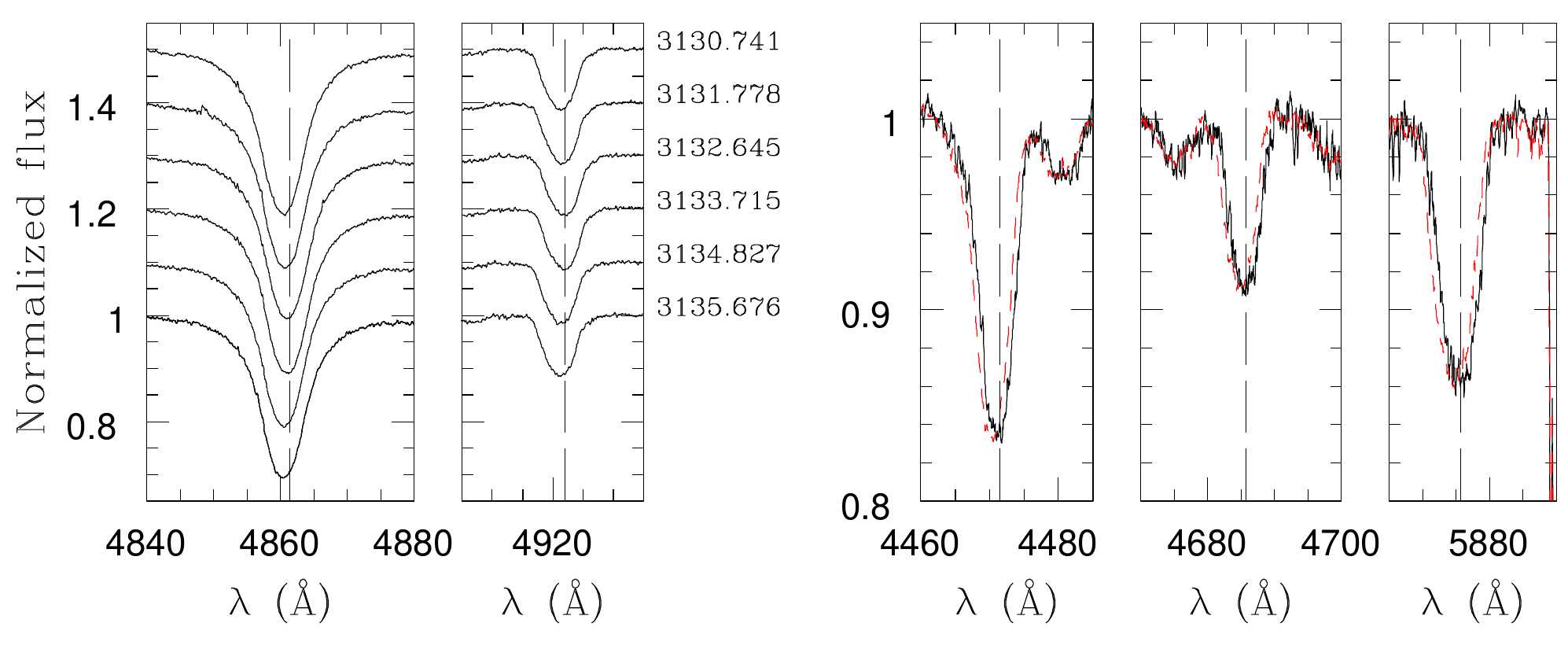}
\caption{{\bf \hd200:} {\it Left panel.} Evolution of the \hbet\ and \hea\,\l4921 lines during six consecutive nights in May 2004. The layout of the figure is similar to the one of Fig.~\ref{fig: hd233a}.  {\it Right panel.} \hea\,\l4471, \heb\,\l4686 and \hea\,\l5876 lines obtained in 2000 (HJD = 2~451~669.900, solid line) and  2004 (HJD = 2~453~135.676, dashed line).  }
\label{fig: hd200_hb}
\end{figure*}
%***********************************************

\subsubsection{\hd200} \label{sect: 200}

\hd200 is a late O-type star located about 3\farcm6 W-SW of \hda. It is quoted as variable by \citet{FF68}, with $V\sim8.35-8.42$. Photoelectric measurements reported in other works since the 1960's are all in the range 8.41-8.44, except the measurement from \citet[ $V=8.37$]{Oja86}. More recent CCD photometry yielded values between 8.31 \citep{BVF99} and 8.42 \citepalias{BL95}. Regarding the spectral types, \citet{SHS69} classified \hd200 as O9.5~III, \citet{Hou78} as O9~III, whereas \citet{LM80} and \citetalias{PHYB90} respectively  preferred O9.5~V(n) and B0~IV classifications. \citetalias{LM83} reported three RV measurements on three consecutive nights, which indicate a velocity change of about 70~\kms\ from one night to the other. The Ca-K corrected velocities of \citetalias{PHYB90}   show a 30~\kms\ range on three measurements spread over 10 days. Finally \citetalias{GM01} claimed to have detected the secondary component signature on 2 out of their 5 spectra. The system should apparently have a period of a few days as the  lines are reported as blended one night and as separated by $\sim180$~\kms\ the following night. 

\begin{figure}
\includegraphics[width=\columnwidth]{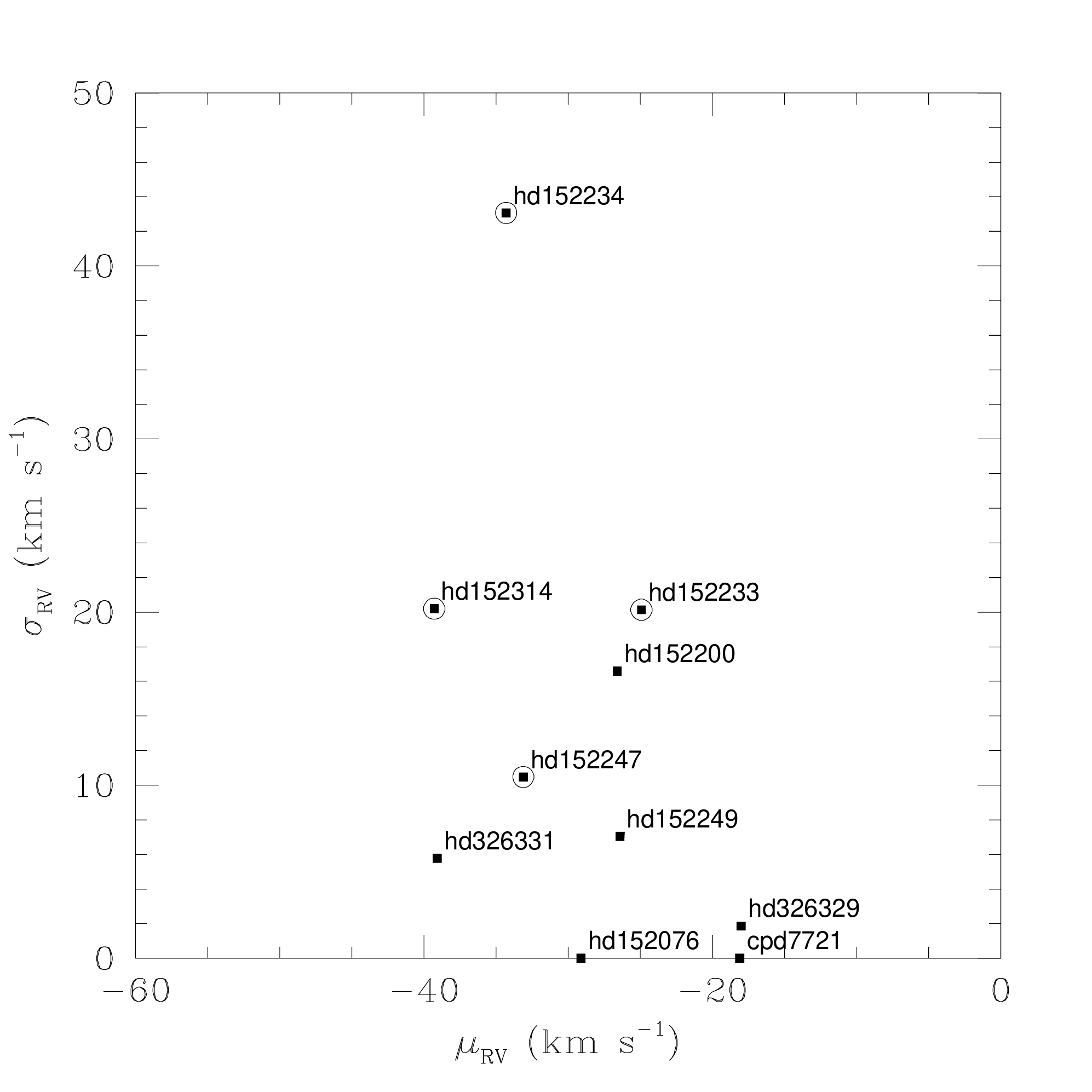}
\caption{Mean RV ($\mu_{\mbox{\tiny{\sc{RV}}}}$) and 1-$\sigma$\ dispersion ($\sigma_{\mbox{\tiny{\sc{RV}}}}$) for the 10 O-type stars/binaries described in Sect.~\ref{ssect: longP} and \ref{ssect: var}. Circles indicate the definite binaries in our sample. This graph is based on the \hea\l5876 line measurements only to preserve the homogeneity of the data set and to avoid biases induced by combining different lines for different objects. $\sigma_{\mbox{\tiny{\sc{RV}}}}$ is arbitrarily set to 0 for the 2 objects with only 1 RV measurement. }
\label{fig: mu-sig}
\end{figure}

From May 1999 until May 2004, we collected 16 FEROS spectra spread over different time scales. The optical spectrum of \hd200 is characterized by the Balmer, \hea\ and \heb\ lines. The few metallic lines typical of late O-type objects are also clearly seen.  The line profiles are variable (Fig.~\ref{fig: hd200_hb}) but no obvious signature of a companion could be found.  The observed variations display a peak-to-peak amplitude of 50\,\kms\ and seem to have a period close to 5 days. The \hbet\ line behaviour (Fig.~\ref{fig: hd200_hb}) is clearly different from what was reported by \citetalias{GM01}. The maximum night-to-night shift is about 30~\kms (to be compared with the $70$~\kms\ shift observed by \citetalias{LM83}).  From our observations, the central value around which the variations occur  seems to have shifted by about 10 to 15\,\kms\ between the years 2002 and 2004. It could however be an apparent effect resulting from a too sparse sampling of the possible $\sim5$-day period. We used the \citet{LK65} and the \citet{HMM85} period search techniques and two aliases were found around $P \sim 4.44$ and 5.56~days. However, both period values yield a large scatter of the RV measurements plotted against a hypothetical phase and it seems difficult to interpret the RV variations in terms of a Keplerian motion. Still, the amplitude of the RV variations compared to the ones of the other O-stars in the cluster (Fig.~\ref{fig: mu-sig}) pinpoints \hd200 as a good binary candidate.

The measured EWs correspond to an O9.7 star with spectral type O9.5 at 1-\s. Conti's criterion ($W''$) clearly indicates a giant class while Mathys's criterion ($W'''$) rather points towards class V. We computed the absolute visual magnitude to be about $M_\mathrm{V}=-4.08$ which is typical of a main sequence O9.7 star. Though binarity would provide the most straightforward explanation to the observed RV variations, we have not been able to satisfactorily fit the data, nor have we been able to detect the companion's signature. More observations are needed to confirm or invalidate the multiplicity status of this object.
% From our data and beyond the night-to-night RV variations,  we find no direct evidence that \hd200 is a spectroscopic binary. \\

\subsubsection{\hd249} \label{sect: 249}

\hd249 is one of the brightest members of the cluster. Its visual magnitude, as quoted since the 1960's, is in the range 6.43-6.51, with the notable exception of \citet{FF68} who reported the star as variable with $V=6.34-6.50$. Quoted spectral types are O9-B0~Ia/b/ab, with the exception of \citetalias{RCB97} who listed O7-O9~V! \hd249 is adopted by \citet{WF90} as an OC9.5~Iab standard.
Several authors indicated RV variations. \citet{Neu30} obtained a $\Delta RV$ range of 46~\kms\ on 3 plates and classified it as a spectroscopic binary. \citet{Str44} and \cite{PHYB90} respectively obtained 15 and 13 measurements on a time-span of 16 and 75 days. Both data sets display a similar range as the one obtained by \citet{Neu30}. However these do not present the smooth variations that could be expected for a binary with a period larger than a couple of days. Later on, \citet{GCM80} and \citet{LMGM88} obtained additional measurements that do not indicate night-to-night variability but, separated by about 5 years, they differ by 20~\kms.

Between 1999 and 2004, we collected  34 spectra. The peak-to-peak dispersion over the 6 years is 20~\kms\ and the night-to-night variability has an amplitude of $\sim10$~\kms. We consider the star to present slight RV changes but clearly, we cannot associate these with a binary nature. The measured EWs unambiguously lead to an O9~I spectral type, thus slightly different from the \citet{WF90} classification. With $V=6.44$ and $B-V= 0.19$, we obtained $M_\mathrm{V}=-6.1$, quite typical of Ib supergiants rather than Iab \citep{Lan92}. With the presence of the \nc\,\ll4634-4641 lines in weak emission, we finally adopt an O9~Ib~((f)) classification.\\

\subsubsection{HD~326331 $\equiv$ \cpd7744} \label{sect: 331}

With $V\sim7.5$ \citepalias{SBL98}, HD~326331 is an O-type star located at 3\arcmin\ East of the cluster center. Together with \cpd7744B ($V\sim10$), it forms a visual pair separated by 7.3\arcsec. If physically related, their revolution period should be over 200\,000 years \citep{MGH98}. Quoted spectral types range from O7~III \citep{LM80, LM83} to O9~III \citepalias{HCB74}, but only give the composite classification. The system was reported as photometrically variable by \citetalias{PHC91}, and \citetalias{HCB74} mentioned an amplitude of 0.3 mag. If we ignore the value $V=7.71$ from  \citet{FF68}, then all the published photometric data since the 1960's  indicate a maximum amplitude of 0.15, with an average value of $7.52\pm0.08$. RV measurements were obtained by \citet{Str44}, \citetalias{PHYB90}, \citetalias{HCB74} and \citetalias{LM83}. The \citetalias{HCB74} data consisted of 13 spectra obtained over 7 consecutive nights. Based only on the \ob\,\l4069 and \sid\,\l4089 lines, they reported smooth velocity variations between +94 and $-124$~\kms\ (mean error of 13~\kms) and then up again. They thus suggested that the star is a binary with a period $P\ga7$~d. They however cautioned that ``the spectrum is very difficult to measure as the lines are weak''. Finally, using one  additional measurement obtained about 7 years later, \citetalias{LM83} published a first SB1 orbital solution with $P=6.24208\pm0.00015$, $e=0.12\pm0.09$ and $K_1=75\pm8$~\kms. The obtained residuals ($\sim$17~\kms) are however very large compared to other orbital solutions published in the same paper.

\begin{figure} \centering
\includegraphics[width=\columnwidth]{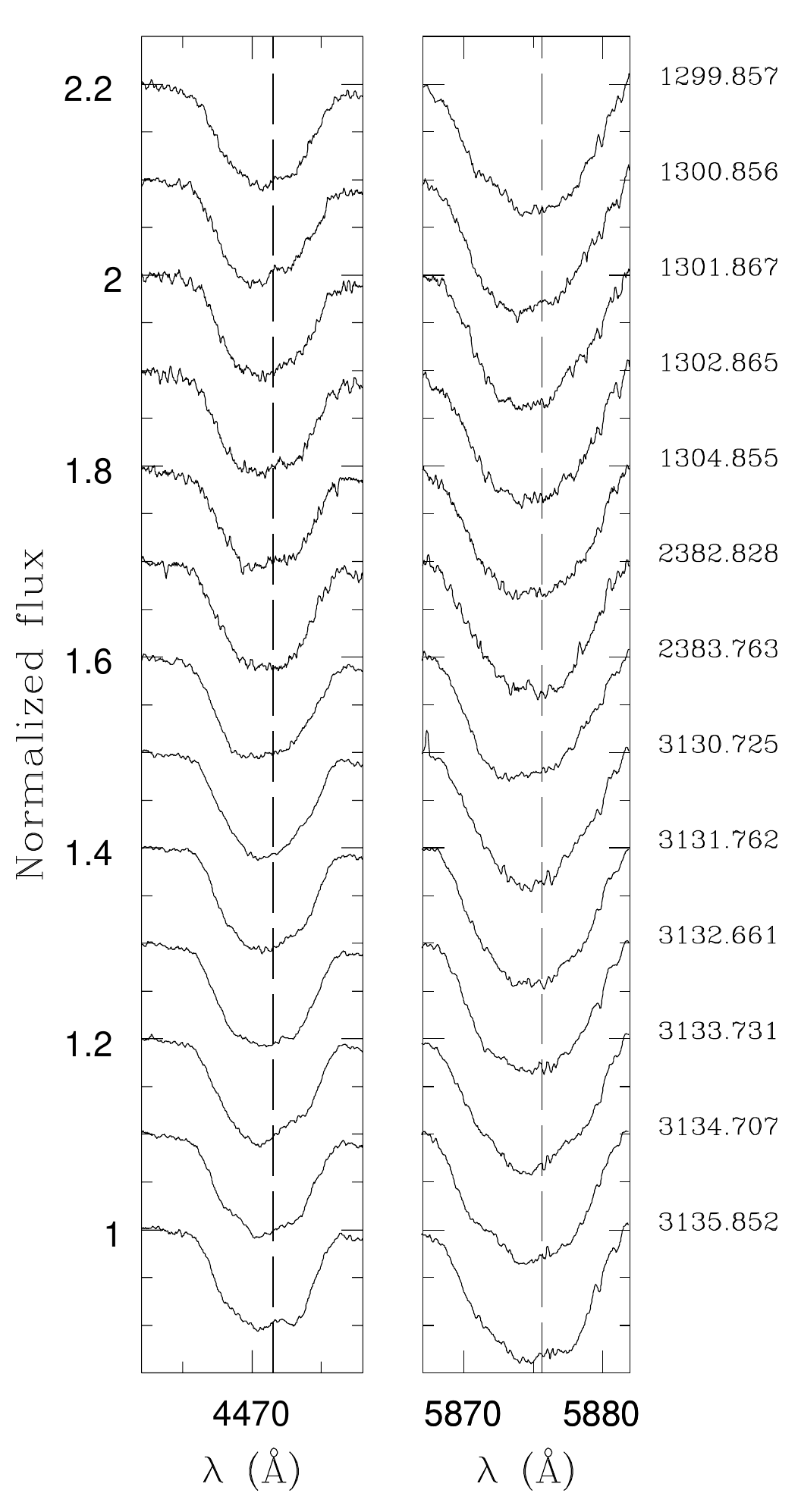}
\caption{{\bf HD~326331:} \hea\,\ll4471, 5876 line profiles as observed from 1999 to 2004 on various consecutive nights.  The layout of the figure is similar to the one of Fig.~\ref{fig: hd233a}.}
\label{fig: hd331a}
\end{figure}

We have sparsely observed HD~326331 since 1998. We obtained 4 spectra of the \hea\,\l4471 region with the CES in May 1998, which revealed very broad lines ($>10$\AA). We then successively acquired FEROS spectra in May 1999 (5), in April 2002 (2) and in May 2004 (6). These were usually separated by one night to cover the proposed 6.4-day orbital period.  The spectrum of HD~326331 presents the clear signature of the usual Balmer, \hea\ and \heb\ lines. Both \heb\,\l4686 and \halph\ profiles show mixed absorption and emission. A faint emission component could also be present in the red wing of \hea\,\l5876. The lines present a flat bottom and their profiles are therefore clearly not Gaussian (Fig.~\ref{fig: hd331a}). The broadness of the lines definitely suggests that  HD~326331 is a rapid rotator and we used the Fourier transform of the line profile \citep[and references therein]{Gray05} to estimate a rotation velocity $v \sin(i) =  310$\,\kms. Then, assuming that the shape of the lines is fully determined by the rotational broadening, we used a rotation profile with $v \sin(i) =  310$\,\kms\ to fit the \hea\ll4471, 5876 and \heb\l4542 lines. The obtained RVs are reported in Table 3 and show 1-$\sigma$ range of about 3 to 9\,\kms. 

%***********************************************
%***********************************************

Though the peak-to-peak dispersion seen in our measurements is about 20~\kms, we  never obtained such highly positive RVs as the ones reported by \citetalias{HCB74}. Still,  Fig.~\ref{fig: hd331a} shows that the spectrum of the star displays obvious profile changes. Though their nature is unclear, it seems difficult to relate them to the presence of a secondary companion and we consider them as intrinsic to the star.

To estimate the spectral type, we measured the EWs of the usual classification  lines. For this peculiar object, we integrated the different line profiles rather than adopting the EWs of the fitted profile. Measured values point towards an O8~I star. However, with the \heb\,\l4686 and \halph\ lines showing mixed absorption and emission, with \nc\,\ll4634-41 weakly in emission and with clearly delineated \hea\,\l4388, the observed spectrum is not that of a supergiant. Furthermore, rapid rotation is rather unlikely for a supergiant. We thus prefer the III((f)) classification. With $V=7.50$ and $B-V=0.18$, the visual absolute magnitude of the object is about $M_\mathrm{V}=-5.2$, which again rules out a supergiant classification. According to \citet{HP89} and \citet{HM84}, this value is rather typical for giant stars. While determining the nature of the observed LPVs requires more follow-up, we can however rule out the 6 or 7 day period previously claimed by \citetalias{LM83} and, until proven to be otherwise, we consider the star as single.\\

%***********************************************
%***********************************************
%%%%%%%%%%%%%%%%%%%%%%%%%%%%%%%%%%%%%%%%%%%%%%%%%%%%%%%%%%%%%%%%%%%

\subsection{Presumably single stars} \label{ssect: sgl}

\subsubsection{\hd076} \label{sect: 076}

Somewhat offset from the cluster core, \hd076 had so far received less attention than the other bright stars. Some confusion seems to exist  about its magnitude and indeed, cross-identifiers reported in the \sbd\ database are clearly erroneous. For example, \sbd\ associates \hd076 with the star Se~309/SBL~350. \hd076 however lies outside the field of investigation of these two studies \citepalias{RCB97, SBL98}.  It is obvious from a comparison with neighbouring objects in DSS images that the $V=10.85$ magnitude reported in \sbd\ is wrong. We adopt in the following $V=8.471$ (Sung 2005, private communication)  which is close to previous determinations of $V=8.5$ \citepalias{PHYB90}, 8.48 \citepalias{PHC91}, 8.47 \citep{SHS69}, 8.50 \citep{FF68}, 8.46 \citep{BBG66} or 8.48 \citep{HW84}. Quoted spectral types are in the range B0~V \citep{MGG53, MWC53} - B0/1~III \citep{SHS69,Hou78}. A couple of RV measurements were also performed by \citet{Str44}, \citet{Wil53} and \citetalias{PHYB90} who obtained a Ca-K corrected RV of about $-$25 to $-$30~\kms. \citet{Bal83} reported this object to be constant in the Johnson B filter on a time scale of 5 hours. 

We acquired one snapshot FEROS spectrum of \hd076 that reveals relatively narrow lines. The \heb\ signature is faint, but clearly seen, and spectral criteria indicate an O9.5~V/III star. With $V=8.471$ and $B-V=0.240$, we obtained $M_\mathrm{V}=-4.4$, an intermediate value between typical absolute magnitudes of O9.5 dwarfs and giants. \hd076 is quite offset from the cluster core and thus might not belong to \ngc.  Instead the star could belong to the \sco\ association, located at the same distance and with a similar age as \ngc. Though its luminosity does not agree perfectly with the spectral type deduced from the spectroscopic criteria, \hd076 is unlikely to be a background object.
Because of the narrow lines, suggesting a lower gravity, and the relatively strong metallic spectrum, we opted for an O9.5~III classification. RV measurements performed on the classification lines plus \mgb\,\l4481 give $\overline{RV}=-30.7\pm2.7$~\kms, in good agreement with previous measurements. We thus consider \hd076 to most probably be a single star, though it is obviously difficult to discard the possibility of a very long period binary. \\

%***********************************************
\begin{figure*}
\includegraphics[width=\textwidth]{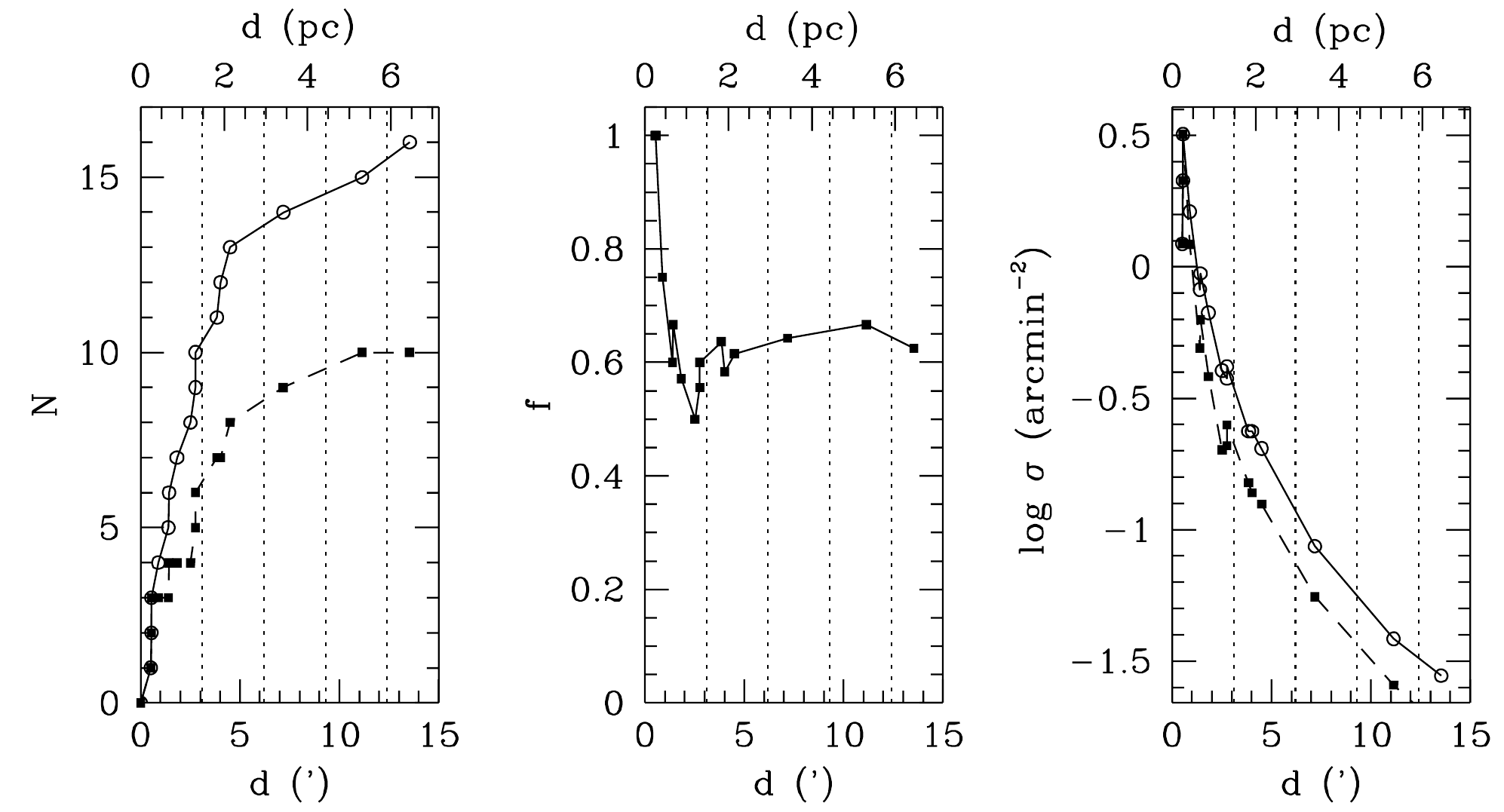}
\caption{  {\bf Left panel:} Cumulative distribution of the number of  O-type objects (solid line and open symbols) and O-type binaries (dashed line and filled symbols) as a function of the distance $d$ from the cluster geometric center. {\bf Middle panel:} Evolution of the binary fraction with $d$. {\bf Right panel:} Integrated O-star number density ($\sigma=N/\pi d^2$) versus $d$. Symbols and lines have the same meanings as in the left panel. In all three panels, vertical dotted lines indicate multiples of the core radius $d_\mathrm{c}=3.1\arcmin\equiv1.5$~pc.}
\label{fig: bf}
\end{figure*}
%****************************************************************

\subsubsection{HD~326329 $\equiv$ \cpd7735} \label{sect: 329}

HD~326329 is located in the core of the cluster. With $V\sim8.8$, it is relatively bright and has thus been observed for a long time. Derived spectral classifications oscillate between O9 and B0 and agree on the main sequence luminosity class.  A relatively large scatter in the reported visual magnitudes led some authors to consider this object as variable. Focusing on CCD data published since the 1990's, we find $V$ in the range 8.71 \citepalias{RCB97} -- 8.81 \citepalias{BL95}. \citetalias{LM83} and \citetalias{PHYB90} respectively obtained 3 and 2 spectra of HD~326329 and found consistent RVs (respectively $\overline{RV}=-30(\pm2)$ and $-30(\pm5)$~\kms). \citetalias{GM01} obtained three additional spectra spread over 6 days. They derived RVs from $-34$ to $-60$~\kms, thus largely scattered and quite different from the cluster systemic velocity, but did not comment on these facts. 

We obtained two FEROS spectra in May 2000 and six additional spectra in May 2004. Within our data set, HD~326329 does not present any significant RV change. For example, we obtained $\overline{RV}=-20.3\pm1.5$~\kms\ for the \hea\,\l4388 line, $\overline{RV}=-18.0\pm1.9$~\kms\ for the \hea\,\l5876 line and $\overline{RV}=-16.5\pm2.2$~\kms\ for the \hea\,\l7065 line. The spectrum of HD~326329 displays the clear signature of an O9.5~V star, with the O9 type at 1-\s.  Finally, with $V=8.76$ and $B-V=0.17$, we obtained $M_\mathrm{V}=-3.8$, which is slightly too faint for an O9.5~V star at the cluster distance.

Though no significant RV shift is found from our data alone, we note that the average RVs reported by different authors are only marginally compatible with our data.  This could suggest a long period modulation but only a long term monitoring of this object could bring an answer to this question. 
\\

\subsubsection{\cpd7721} \label{sect: 7721}

Together with \cpd7721s ($V=8.90$), \cpd7721 ($V=8.72$) is a visual double star with components separated by $\sim$5\farcs8 and corresponding to SBL 350 and 351 or BVF 12 and 27.  Two spectra were obtained by \citetalias{PHYB90} indicating $RV=-23$ and $-28$\,\kms (mean error $\sim$10~\kms). \citetalias{GM01} also acquired two spectra of the bright component and quoted $RV=-26.8$ and $-33.0$~\kms. They assigned it a spectral type O9.5~V. 

In May 2004, we obtained one FEROS spectrum of \cpd7721 and another of \cpd7721s. Each of the two spectra show a single spectral signature. Consequently, we consider the two stars as apparently single. \cpd7721 is an O9~V star with an average RV of $-25.5$~\kms, although we observed a large scatter (from $-36$ to $-13$~\kms) according to the various lines measured. With no \heb\,\l4542 line and a faint \heb\,\l4686 line, \cpd7721s is clearly an early B-type star. The strengths of the \sic\,\l4552 and \sid\,\l4089 lines are very similar. Comparing with the spectral atlas of \citet{WF90}, we finally adopt a B1~V class. With $V=8.709$ (resp. $V_\mathrm{s}=9.892$) and $(B-V)=0.168$ (resp. $(B-V)_\mathrm{s}=0.178$), the visual magnitude of  \cpd7721 (resp.  \cpd7721s) is about $M_\mathrm{V}=-5.2$ (resp. $M_\mathrm{V, s}=-2.6$). Both stars are a few tenths of a magnitude too faint if located in the cluster core. Alternatively,  slightly later spectral types would yield a much better agreement.
\\

\section[]{Discussion} \label{sect: disc}

\subsection{Minimum SB fraction}

As mentioned earlier, and taking into account \wr,  16 objects containing at least one O-type star are located within 15\arcmin\ from the cluster center. As shown in Figs.~\ref{fig: Oiden} and \ref{fig: bf}, these stars are unevenly spread across the field of view. Most of the objects are located in the vicinity of the cluster center: ten of them lie within one cluster core radius ($d_\mathrm{c}=3.1'$), while 13 of them are to be found within $1.5\times d_\mathrm{c}$. Among these 16 objects, 6 are close SB2 binaries with periods below 10 days, 4 are longer period systems, 3 stars display line profile variations (LPVs) and 3 stars show  constant RV and constant line profile. 

Given the evidence presented in Sect.~\ref{sect: objects}, no doubt remains about the multiple nature of the 6 short-period and of the 4 longer period systems, though the orbital properties of the latter ones await tighter constraints. This yields a firm lower limit on the  fraction of binaries in \ngc\ of  $f_\mathrm{min}\sim0.63$. 

In Sect.~\ref{sect: objects}, we further noted that other objects were presenting hints of binarity, among which \hd200 and HD~326329 are probably the two best candidates. If the multiplicity of these objects is confirmed, this would significantly increase the binary fraction compared to the lower limit mentioned above. Yet we consider that, so far, there is not enough observational evidence to support such an assertion. 

\subsection{Statistical significance of the results}

While the previous paragraph established that in NGC 6231 the true O-type binary fraction $f$ is located between 0.63 and 1.00, the present results have implications beyond the sole case of NGC 6231 and allow us to put some interesting constraints on the parameters of the underlying distribution of single vs. binary stars. 

Let us adopt a binomial distribution B($n$, $p$), with $n$ the number of trials and $p$ the probability of realisation, to describe the current experiment (i.e. counting multiple stars in a sample of $n$ objects). Given the current realisation of the experiment (10 multiple stars among 16 objects), one can compute that the probability of realisation $p$ of the parent distribution has to be larger (at the 0.01 significance level) than 0.37. In the hypothesis that some binaries have remained undetected despite our efforts, the parent distribution parameter $p$ would be drawn towards larger values, so that the lower limit  $p>0.37$ still holds at the 1\% significance level (or better).

\subsection{Orbital parameters}
\begin{figure}
\includegraphics[width=\columnwidth]{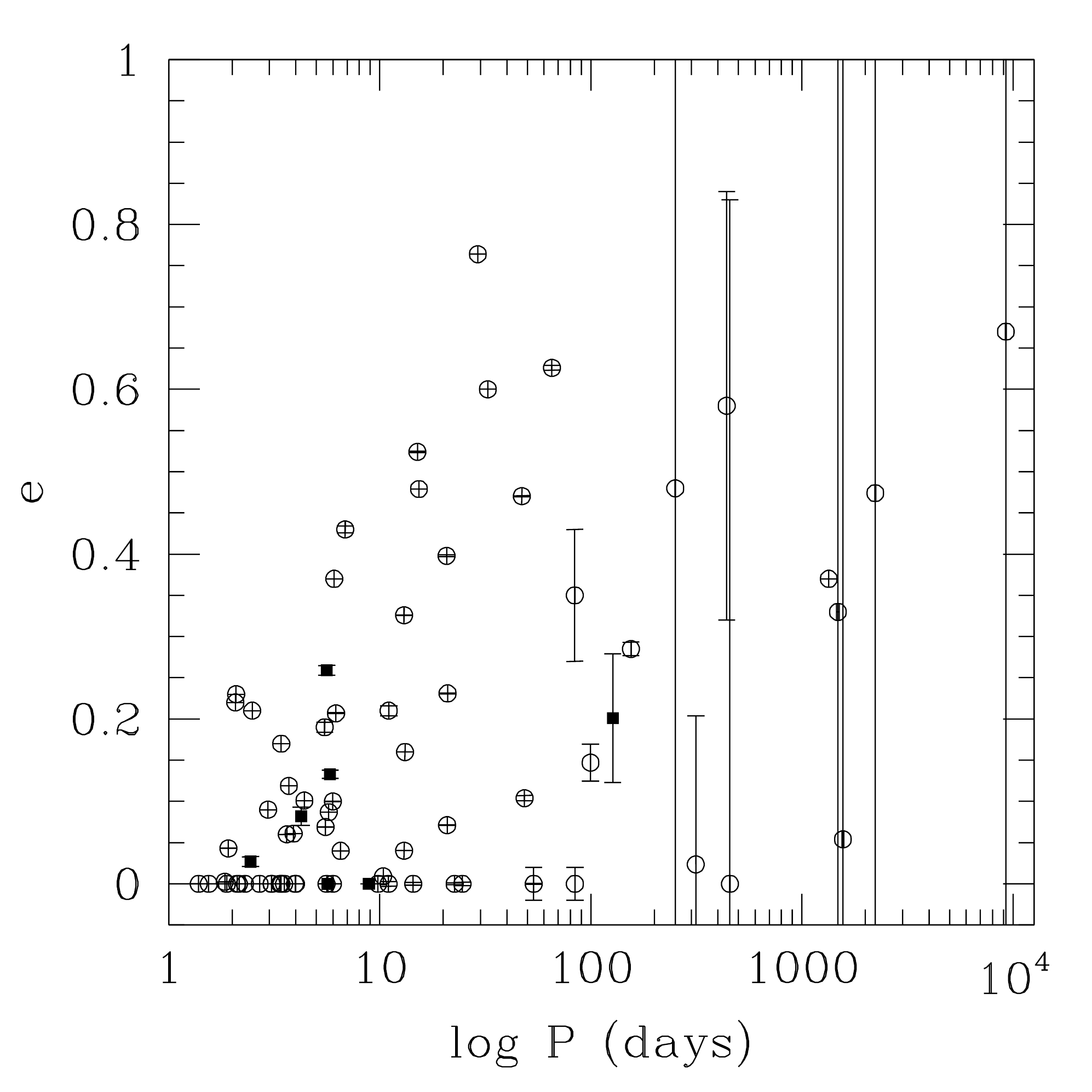}
\caption{ Period -- eccentricity distribution for the O-type SB systems listed in the 9th Spectroscopic Binary catalogue \citep{PTB04}. Filled symbols indicate the location of the \ngc\ systems.}
\label{fig: pe}
\end{figure}

The orbital parameters that we have derived for the short period binaries in \ngc\ (see Table~\ref{tab: shortP}) also reveal a significantly different picture than previously accepted. 
In particular, while \citetalias{GM01} proposed eccentricities up to 0.6 for the short period binaries, we discard these values as resulting either from erroneous Julian dates that pull the eccentricities towards larger values \citep[see e.g.\,][]{SHRG03} or from ill-constrained periods biasing the orbital fit. From our results, none of the short period binary stars has an eccentricity larger than 0.3. As a consequence, the \ngc\ binaries are now in line with the period-eccentricity distribution of other O-type binary objects (Fig.~\ref{fig: pe}).  This has significant implications on the dynamical evolution of the O-type stars as no other binaries with a high eccentricity and a very short period are known so far. 

\begin{figure}
\includegraphics[width=\columnwidth]{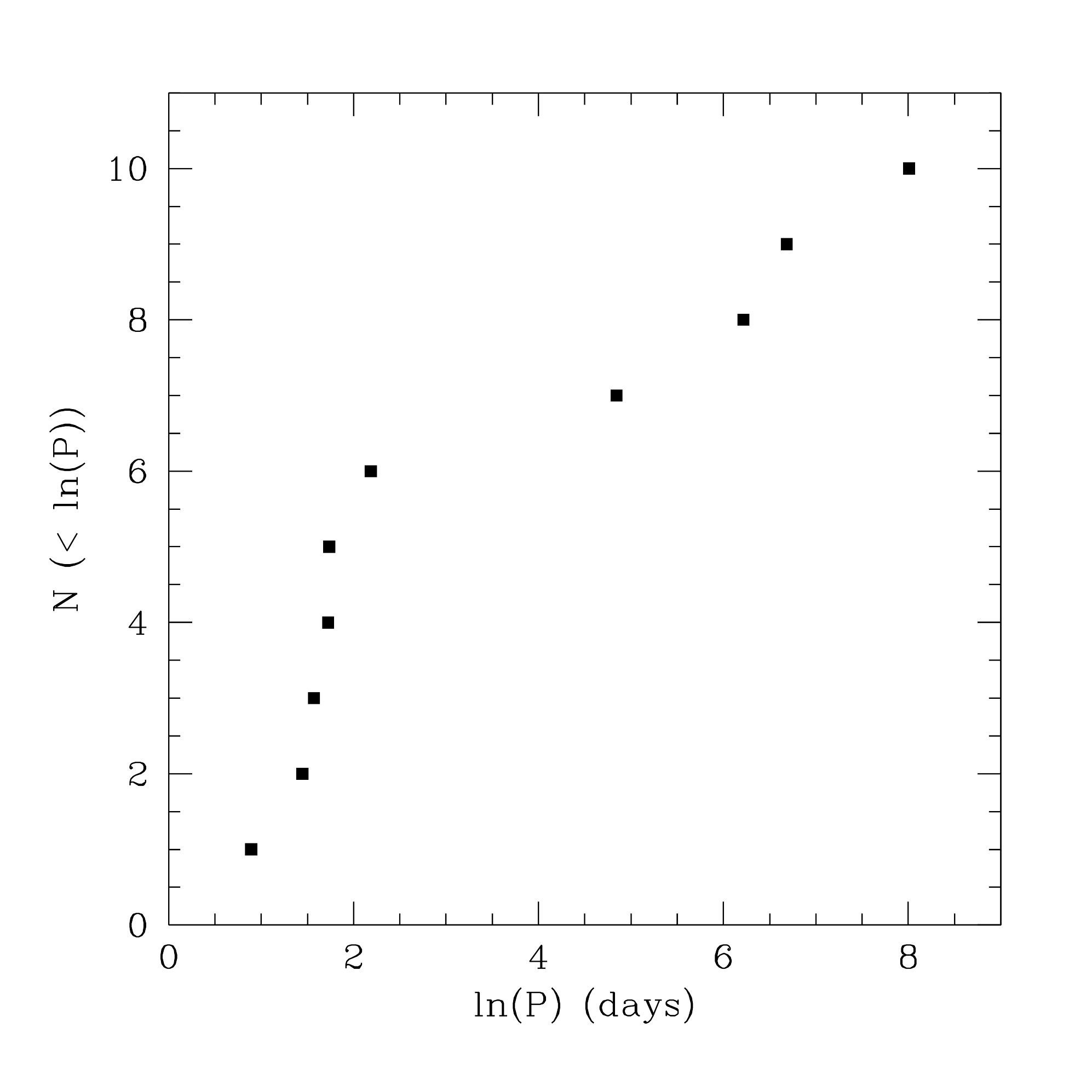}
\caption{ Cumulative number of binary systems, in \ngc, as  a function of the natural logarithm of the orbital period. Period values above $\ln(P)>4$ are very preliminary, though the order of magnitude should be correct. }
\label{fig: lnp}
\end{figure}

 We also note that the short-period systems are significantly more abundant than longer period systems, even compared to \"{O}pik's law which states
 that the distribution of the periods (or equivalently, given Kepler's third law, of the separations) should be flat in the logarithmic space.
Indeed, this would translate itself into a linear dependence of the cumulative distribution of the systems, expressed versus the logarithm of the period while Fig.~\ref{fig: lnp} shows that, in \ngc, this is not the case. Even if all the non-binaries in the cluster were actually undetected long-period binaries, this fact would remain, so that observational biases cannot be invoked.  It is however difficult to decide whether this is a signature of the formation process or of the early evolution history. 

%****************************************************************
%****************************************************************

%****************************************************************
%****************************************************************

\subsection{Mass-ratios and companion IMF}

Not only is the present study based on a complete and homogeneous population of O-type stars, but we emphasize that the signature of all the secondary companions could be unveiled in the detected binaries, yielding an unprecedented view of the binary population parameters. As a matter of fact, all the 10 detected O-type binaries have an OB companion, suggesting luminosity ratios ($L_1/L_2$) between 1 and 10, and mass ratios ($m_1/m_2$) between 1 and 3. 

Of course the present study is biased towards the detection of relatively massive companions. Nonetheless, taking advantage of the completeness of the population, one can already make the following strong statement. Even if the 6 remaining presumably single stars are undetected binaries (low mass or faint companions, pole-on view, etc.), a minimum of two thirds of the companions have a mass above 5\msol. Even considering the possibility of hierarchical multiple systems, the present data set definitely rules out the possibility that the companions in NGC\,6231 are randomly drawn from a Salpeter initial-mass function (IMF). Indeed, if this were the case, one would expect, for each O+O system, to encounter  9, 13, 19, 22, over 100 and a few hundreds of O type binaries with a companion of spectral type B, A, F, G, K and M respectively \citep{WK07}. Considering that, in the present work, one would probably detect O and B companions only, and assuming a 100\% binary fraction, a rule of thumb would predict a binary detection rate of 2\% only, thus about 0.3 binary given the 16-star sample. Despite the limited size of the sample, the null hypothesis that the companions of an O-star are randomly drawn from a Salpeter IMF can be rejected at the 5-$\sigma$ level\footnote{We note that using a standard Kroupa IMF instead of a Salpeter IMF does not significantly change the present result.}.
 
%For completeness, we note that the companion is usually define as the less massive star in the system. Therefore, even if the stars are randomly paired using an underlying Salpeter IMF, the companion IMF will be steeper than a classical IMF \citep{MaZ01, WK07}. In the present case, we actually observe a strong flattening, suggesting that this possible steepening is not the cause of the observed deviation.

\subsection{Visual companions}\label{ssect: vis_comp}

The speckle interferometric campaign of \citet{MGH98} included the brightest objects ($V<8$) of the cluster and revealed that \hd233, \hd234, \hd248 and, possibly, \hd249\footnote{ \citeauthor{MGH98} suggested that \hd249 could have a binary companion at a separation of 0\farcs06, but they emphasized that this result needs to be confirmed.}   have one or several visual companions with a sub-arcsec separation.   Beside these results, direct imaging indicates that \cpd7721s is located at 5\farcs8 from \cpd7721, that HD~326329 has a visual companion at 7\farcs3 and that HD~152076 is not distant by more than 8\arcsec\ from two other stars. Only HD~152200 remains seemingly alone in a 20\arcsec\ radius. So far, however, it is almost impossible to state whether these visual pairs are physically linked, or arise by line-of-sight coincidence in such a crowded field.\\
Using only stars brighter than $V=17$, one can estimate a density of about 10 stars per arcmin$^2$ at the cluster center and about half this value at $1.5\times d_\mathrm{c}$ \citep{SGR06}. Rough statistics  thus suggest that companions separated by less than 1 and 2\arcsec\ respectively (according to whether they are located in the very core or in the outer part of the cluster) are unlikely to arise by chance and are thus likely bound (at the 0.01 significance level). According to this simplistic criterion, only \hd234, \hd248 and, possibly, \hd249 have a physical faint companion with putative revolution periods between 150 and 4900~yr \citep{MGH98}. If verified, this would indicate that at least one third of the SB systems are  hierarchical triple systems, yielding a number of companions per massive star close to 0.8. Such a value remains still far below the 1.4 companion  fraction  derived  by \citet{PWZ01} for the O- and B-type stars in the Orion Nebula Cluster. As a matter of comparison,  either a significant fraction of close and unresolved companions, with a physical separation roughly in the range 2-200 A.U.,  are still missing, or one would need to consider all the known visual companions up to 10\arcsec\ to reach a similar rate for \ngc.

\subsection{Physical constraints for massive star formation}

Because of its spectroscopic approach, the present study is clearly biased toward the detection of SB systems with an OB star companion and only provides a lower limit on the true binary fraction.  Even considering the work of \citet{MGH98}, about 2 orders of magnitude in semi-major axis (i.e. between $\sim~2$ and 200~A.U.) are still not sampled. Investigations of this range would require 10 m/s RV accuracy or milli-arcsec spatial resolution, both of which, as far as O-type stars are concerned,  challenge the current instrumentation limits.  Still the present study provides strong constraints for massive star formation: 

\begin{enumerate}

\item[-] At least 60\% of the O-type objects in \ngc\ are SB systems. Over 70\% of any O-type stars in the cluster are found to belong to an O+O or an O+B system. 
\item[-] If the properties of the O star population in \ngc\ are representative of the O star population in general, the expected binary fraction of the parent distribution should, at the 0.01 significance level, be larger than 0.37. 
\item[-] Concerning the 10 SB binaries in our sample, half of them have an O-type companion, the other half having a B0-3 companion. Therefore, in all the detected SB binaries in the present sample, the luminosity ratio is about or less than 10 to 1 while the mass ratio is about or less than 3 to 1. 
\item[-] Even in the extreme case of a 100\% binary content, it is very unlikely that the companion of an O-type star would be randomly drawn from a standard IMF. 
\item[-] As a corollary, the previous statement shows that the massive star formation and/or early evolution tend to produce a large number of double objects with components having masses of the same order of magnitude.
\item[-] 60\% of the detected SB systems (thus about one third of the O-type objects) are tight binaries with periods between 2.4 and 9 days, and with zero or small eccentricities.  Compared to \"Opik's law, short period systems are overabundant, even in the extreme case of $f=1.0$. 
%\item[-] The number of O-type objects encountered increases almost linearly with the distance from the cluster center up to $\sim1.5d_\mathrm{c}$, which means that the number density decreases with $d^{-2}$ (Fig.~\ref{fig: bf}). 
\item[-] No significant difference in the spatial distribution of the binary fraction can be found across the cluster (Fig.~\ref{fig: bf}).
\end{enumerate}

%_____________ CONCLUSIONS______________________________________________

\section{Conclusions} \label{sect: ccl}

Based on a set of high-resolution high SNR spectra of the O-type objects in \ngc, covering various time scales from a few hours to several years, we have revised their physical and orbital status, paying  particular attention to their possible multiplicity. The present fraction of binaries in \ngc\ (around 63\%) is significantly reduced compared to the value of 79\%  obtained by \citetalias{GM01}. %This mainly results from the revision of the status of two objects in our sample, \hd200 and HD~326331 which were previously reported as binaries.  

We note that the revision of the binary fraction in \ngc\ puts the cluster in a less extreme position compared to other rich (in terms of the number of O-type stars) open clusters. Considering also that \citet{DBRM06} could not reproduce the high binary fraction ($f\sim0.80$) found by \citetalias{GM01} for the cluster IC~1805 either,  the anti-correlation proposed by \citetalias{GM01} between the cluster density and its massive star binary fraction, remains to be confirmed.

While detailed studied of other open clusters are needed, the present approach already offers the possibility to quantitatively test the null hypothesis that the O-type SB populations of different clusters are drawn from the same distribution. Indeed, the rejection of this hypothesis is a pre-requisite to the suggested link between the binary fraction and the cluster properties. 

Beyond the binary fraction, we also conclude that, in \ngc, the massive stars are not randomly paired from a standard underlying IMF. Instead, there is a strong bias towards the formation of O+OB systems. Among the latter, about half of them are close binaries with a period of a few days. Finally, none of the short period binaries has an eccentricity larger than 0.3. Compared to the periods and eccentricities found by \citetalias{GM01}, this has strong implications on the dynamical evolution of such systems.   

The present results  outline the limitations of previous estimates of the properties of the O-type population in young open clusters. It also emphasizes the need for extensive studies to accurately constrain, among other properties, the binary fraction of various stellar populations and, particularly, of early-type stars whose formation and evolution are still not firmly understood.

%Beyond the large binary fraction observed, we show that the O-type binary population is strongly biased towards low mass-ratio binaries and, as a consequence, that the companion can not been randomly draw for a classical IMF. These figures provide strong constraints for the formation and early evolution of O-type stars and binaries. Indeed, even if \ngc\ is a single cluster, theories have now to be able, choosing appropriate initial conditions corresponding to \ngc, to reproduce the main properties of the cluster massive star population: a high binary fraction, among which a high rate of O+OB system. 

%A forthcoming paper in the series will present new data for two other rich young open clusters : IC~2944 and NGC~6611. The aim pursued will remain the same : to study the possible link between the environment and the massive star population, and to provide strong observational constraints to formation and early evolution theories.

%CCl: pap3: In conclusions, the anti-correlation between the cluster density and its massive star binary fraction, in rich cluster, remains to be confirmed. 

%_____________ ACKNOWLEDGMENTS __________________________________________

\section*{Acknowledgments}
%It is a pleasure to thank Michael Sterzik and Hans Zinnecker for helpful discussions.
The authors are grateful to Jean-Pierre Swings and to Michael West for helpful comments on the manuscript. The Li\`ege team aknowledges support from the FNRS (Belgium). This work made use of the SIMBAD and WEBDA databases and of the Vizier catalogue access tool (CDS, Strasbourg, France). It further relies on data taken at the La Silla-Paranal Observatory under program IDs 061.D-0502, 063.H-0061, 063.H-0093, 065.H-0265, 067.D-0059, 068.D-0095, 069.D-0381, 071.D-0369 and 073.D-0609. We thank the ESO staff for efficient support during both visitor and service mode programs.  

%_____________ BIBLIOGRAPHY _____________________________________________

\bibliographystyle{mn2e}
\bibliography{/home/hsana/disk-externe/LIEGE_PAPERS/XMM_CAT_PAPER4pdf/ngc6231_Xcat}

\bsp

\label{lastpage}

\end{document}